\documentclass{elsart4-1}

 \usepackage{graphics}
 \usepackage{graphicx}
\usepackage{amssymb}

\usepackage[english,francais]{babel}

\newtheorem{e-proposition}[theorem]{Proposition}

\newtheorem{e-definition}[theorem]{Definition\rm}

\renewcommand{\theequation}{\arabic{equation}}
\def\og{\leavevmode\raise.3ex\hbox{$\scriptscriptstyle\langle\!\langle$~}}
\def\fg{\leavevmode\raise.3ex\hbox{~$\!\scriptscriptstyle\,\rangle\!\rangle$}}

\begin{document}
% Select a primary header Physics or Astrophysics
% You can place after the header (classification), if you know it.

\centerline{Physics}
\begin{frontmatter}


% \author{Name\thanksref{label2}}
\selectlanguage{english}
\title{Novel magnetic order in pseudogap state of high Tc copper oxides superconductors}

% If all authors are at the same address, the [label1] can be suppressed

\selectlanguage{english}
\author[authorlabel1]{Philippe Bourges},
\ead{philippe.bourges@cea.fr}
\author[authorlabel1]{Yvan Sidis}
\ead{yvan.sidis@cea.fr}

\address[authorlabel1]{Laboratoire L\'eon Brillouin-Orph\'ee, CEA-Saclay, 91191 Gif-sur-Yvette, FRANCE}
% \address[authorlabel2]{Laboratoire Léon Brillouin-Orphée, CEA-Saclay, 91191 Gif-sur-Yvette, FRANCE}

% If your know the dates of reception, and acceptation you can put them now;
%    idem the name of the person presenting your article

\medskip
\begin{center}
{\small Received *****; accepted after revision +++++}
\end{center}

\begin{abstract}
One of the leading issues in high-$\rm T_c$ copper oxide superconductors is the origin of the pseudogap phase in the underdoped regime of their phase diagram. Using polarized neutron diffraction, a novel magnetic order has been identified as an hidden order parameter of the pseudogap as the transition temperature corresponds to what is expected for the pseudogap. The observed magnetic order preserves translational symmetry as predicted for orbital moments in the circulating current theory. Being now reported in three different cuprates familles, it appears as a universal phenomenon whatever the crystal structure (with single CuO$_2$ layer or bilayer per unit cell). To date, it is the first direct evidence of an hidden order parameter characterizing the pseudogap phase of high-$\rm T_c$ cuprates.
\\
{\it To cite this article: P. Bourges and Y. Sidis, C. R. Physique xx (2010).}

\vskip 0.5\baselineskip

\selectlanguage{francais}
\noindent{\bf R\'esum\'e}
\vskip 0.5\baselineskip
\noindent
{\bf Ordre magn\'etique de l'\'etat pseudogap des  oxydes de cuivre supraconducteurs \`a haute temp\'erature critique}
Un des probl\`emes majeurs des oxydes de cuivre supraconducteurs \`a haute temp\'erature critique est l'origine de la phase pseudogap dans l'\'etat sous-dop\'e. En utilisant des mesures de diffraction de neutrons polaris\'es, nous avons identifi\'e un ordre magn\'etique cach\'e qui appara\^it \`a la temp\'erature attendue de cet \'etat de pseudogap. Cet ordre  magn\'etique respecte la sym\'etrie de translation du r\'eseau comme cela a \'et\'e pr\'edit dans la th\'eorie des boucles de courants circulants. Nous avons g\'en\'eralis\'e cette d\'ecouverte dans trois familles distinctes de cuprates. Cette mesure est la premi\`ere preuve exp\'erimentale directe d'un param\`etre d'ordre universel de l'\'etat de pseudogap des oxydes de cuivre supraconducteurs.
\\
{\it Pour citer cet article~:  P. Bourges et Y. Sidis, C. R. Physique xx (2010).}

%Now keywords/mots-clÈs
 \vskip 0.5\baselineskip
\noindent{\small{\it Mots-cl\'es~:} diffraction de Neutron, supraconducteurs à haute temp\'erature critique, phase de pseudogap}}
\end{abstract}
\end{frontmatter}

% now the Version franÁaise abrÈgÈe, if it exists
%\selectlanguage{francais}
%\section*{Version fran\c{c}aise abr\'eg\'ee}
% Text of your Version franÁaise abrÈgÈe here

\selectlanguage{english}
\newpage 

\section{Introduction}

The origin of high-$\rm T_c$ superconductivity in copper oxide materials is still hotly debated more than twenty years after its discovery. 
On the one hand, conventional phonon-mediated superconductivity has been advocated to explain anomalies in electronic spectroscopies although the electron-phonon coupling seems insufficient to explain the high value of the critical temperature. On the other hand, the antiferromagnetic (AF) spin fluctuations observed in these strongly electronic correlated materials  could also lead to an unconventional superconducting (SC) pairing mechanism, where the conventional electron-phonon coupling would be replaced by a spin-fermion coupling. For instance, unconventional AF excitations, the so-called {\it resonance modes}, have been reported in the SC state of most of cuprates families \cite{cras}. Assuming a rather large spin-fermion coupling ($\sim$1 eV), these collective AF excitations may account for several anomalies in charge excitation spectra and for the angular momentum dependence of the SC gap (see for instance Ref.~\cite{Onufrieva-PRL09}). However, the spin fluctuation mediated pairing scenario has to face a serious problem:  high-$\rm T_c$ superconductivity survives even when the AF spin fluctuation spectral weight becomes strongly reduced. If certain aspects of the physics of cuprates can be understood using either phonons or AF spin fluctuations, the mystery of high-$\rm T_c$ superconductivity in cuprates remains unsolved.

%----------------------------------------------------------------------------------------------
\begin{figure}[b]
\centering
\includegraphics[width=10cm,angle=0]{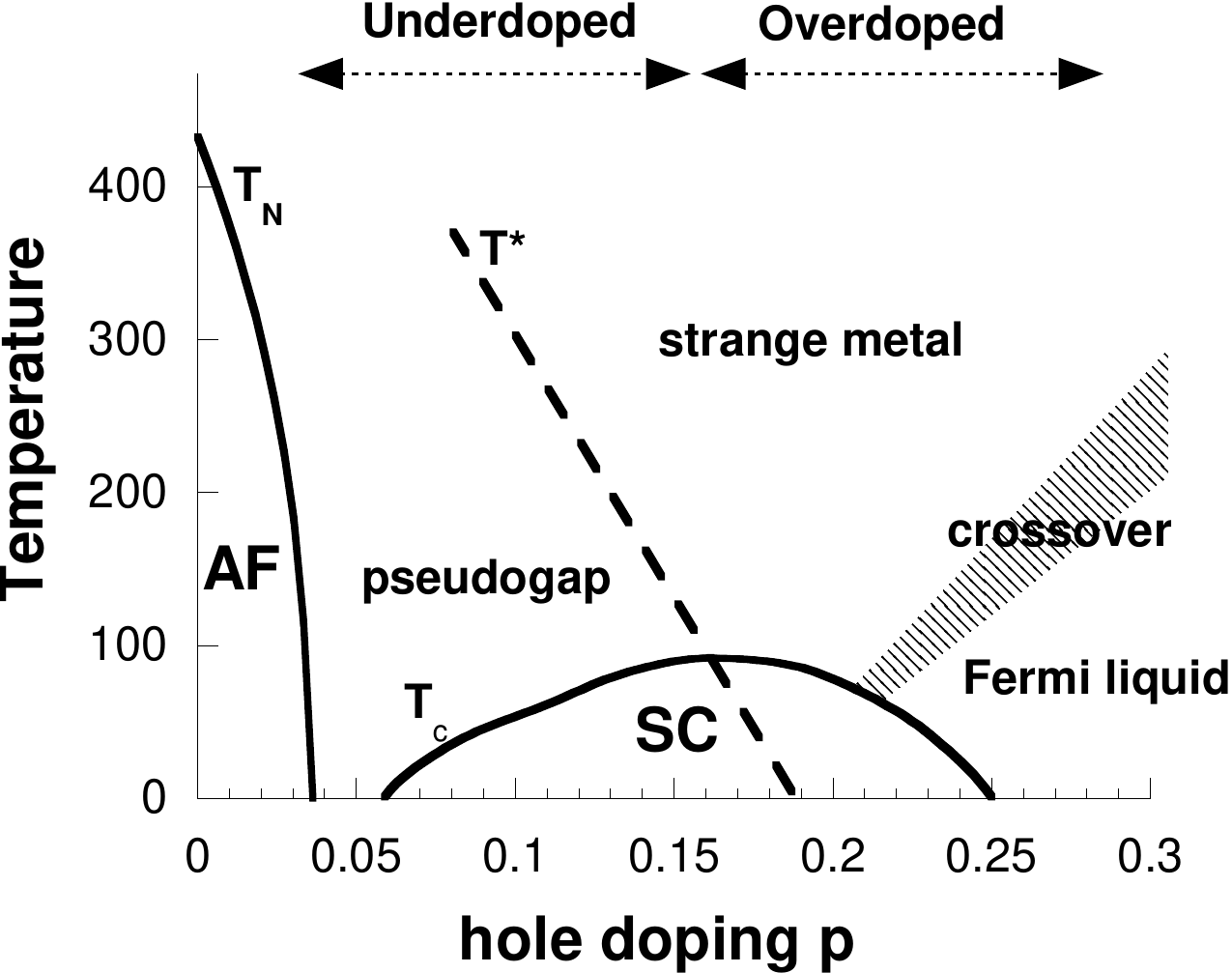}
\caption {Generic phase diagram of high temperature superconducting copper oxides as a function of hole doping (p).
In the Mott insulation state, the antiferromagnetic phase develops below the N\'eel temperature ($\rm T_N$). Upon doping, the system becomes metallic and superconducting (SC) below the superconducting critical temperature ($\rm T_c$). In the normal state, the lost of spectral weight on certain portions of the Fermi surface highlights the opening of the mysterious pseudogap phase below the temperature T*.  
}
\label{Fig1}
\end{figure}
%----------------------------------------------------------------------------------------------

All the high-$\rm T_c$ SC cuprates share a common crystallographic structure: They are layered materials characterized by the stacking of $\rm CuO_2$ planes. These planes are paved with squared $\rm CuO_2$ plaquettes. The charge density in $\rm CuO_2$ planes can be tuned  using either electron or hole doping. All the hole doped high-$\rm T_c$ SC cuprates exhibit the same remarkable phase diagram (Fig.~\ref{Fig1}). 
These compounds are AF Mott insulators at zero doping. The AF state is quickly destroyed once a small amount of doped holes is introduced in the $\rm CuO_2$ planes. Increasing the hole doping, the system becomes metallic and superconducting below the critical temperature $\rm T_c$. At optimal doping (p$\sim$0.16 holes /Cu), $\rm T_c$ reaches its maximum value. Two distinct regimes develop on both sides of the optimal doping. In the overdoped regime when increasing the hole doping, the electronic properties in the normal state can be described using a moderately correlated Fermi liquid picture. At variance, in the underdoped regime when reducing the hole doping, the materials behave as a strongly correlated metals and the standard Fermi liquid picture fails to account for their unconventional electronic properties. In particular, the underdoped regime is dominated by a phase with highly unusual physical properties where magnetic, transport and thermodynamic measurements point towards a diminution of the electronic density of states below a temperature T* \cite{revue,timusk,tallon} although the system remains metallic. As the density of states remains non-zero at the lowest temperature, this phase has been named as a pseudogap phase since its first evidence\cite{Alloul-1989}. However, below T*, a q-dependent gap opens in the single-particle excitation spectrum measured in angle-resolved photoemission (ARPES) experiments with a maximum at the wave vector M=($\pi$,0) and symmetry related points of the Brillouin zone. Other spectroscopies\cite{timusk} spanning from scanning tunneling microscopy (STM), electronic Raman spectroscopy and optical conductivity  reported  anomalies at similar energy.  The understanding of the microscopic origin of the pseudogap phase and of its interplay with unconventional $d$-wave superconductivity has become one of the major challenges to overcome in order to crack the mystery of  superconducting cuprates.

Two major classes of theoretical models attempt to explain 
the pseudogap state. It has been  at first proposed that the pseudogap phenomenon can be described as a precursor of the superconducting $d$-wave gap, but with no phase coherence, which would occur only below T$_c$ \cite{preformedpairs}. This preformed pairs scenario has largely been favoured over the years in relation to the doping dependence of the SC gap which does not follow $\rm T_c$ (as it is expected in the conventional BCS theory of superconductivity) but rather T*. This has been discussed in conjunction of many experimental techniques such as ARPES, STM, electronic Raman scattering experiments. However, due to lacking experimental evidence of such preformed pairs\cite{bergeal}, this approach is questioned.  

Many other theories attribute the pseudogap origin to the proximity of a competing state, but there is a wide disagreement about the nature of this state. In certain theoretical models, the pseudogap state is a long range ordered phase, with a well defined order parameter and a related broken symmetry \cite{cmv97,simon,cmv06,ddw}. In this scenario, the ordering temperature T* and the order parameter should vanish at a quantum critical point (QCP) located behind the SC dome, close to optimal doping (see dashed line in Fig. \ref{Fig1}). A large number of experimental properties ranging between transport, thermodynamic and magnetic \cite{tallon} point towards the existence of a QCP for a doping level of  p=0.19. The order parameter may involve charge and spin density waves or charge currents flowing around or inside CuO$_2$ plaquettes \cite{cmv97,simon,cmv06,ddw}. Interestingly, while the low-doping phase may compete with superconductivity, the fluctuations associated with the broken symmetry are expected to be rather strong around the QCP and could play the role of pairing glue leading to high temperature superconductivity. This appealing scenario has nevertheless to face two major experimental facts: there is no clear jump of the specific heat at T*, as expected for many phase transitions (but not all) and there is no indication that the translation symmetry of the lattice (TSL) is broken in the pseudogap state. 

Alternatively, other theoretical approaches have been developed still on the basis of a competing electronic instability resulting from strong electronic interaction, but in these approaches the pseudogap state is no longer an ordered phase \cite{stripes,vojta,Onufrieva-PRL99,Chubukov-PRB10,Sachdev-review}. It is a disordered phase dominated by the fluctuations associated with a competing electronic instability and T* is only a crossover of dynamical properties. Nevertheless, a true ordered state could be stabilized at low temperature at the expense of superconductivity when applying an external perturbation (external magnetic field, disorder, structural distortion, etc...). 
Such an approach is well illustrated by the stripes model where charges self-organize  to form lines, separated AF domains in anti-phase \cite{stripes}. When 1D charge stripes appear, but still fluctuate, the C4 rotational symmetry is first broken. When fluctuating stripes are pinned down on the lattice or defects, they become static and the resulting charge and spin order ultimately breaks the TSL. In ${\rm La_{2-x}Sr_xCuO_4}$ (LSCO), incommensurate spin fluctuations are usually interpreted in terms of fluctuating stripes,  that can be pinned down at low temperature by disorder or stabilized under an external magnetic field \cite{Tranquada-review}. A static stripe phase has been further reported in ${\rm La_{2-x}Ba_xCuO_4}$ or in ${\rm (La,Nd)_{2-x}Sr_xCuO_4}$ around the critical hole doping p=1/8.
In addition, in strongly underdoped ${\rm YBa_2Cu_3O_{6.45}}$ (p=0.085) \cite{hinkov} as well as for similar dopings \cite{haug1}, anisotropic incommensurate spin excitations are observed at low temperature, suggesting that the C4 rotational invariance of the system is spontaneously broken below $\sim$ 150 K. These fluctuations further 
freeze at very low temperature, yielding a glassy short range spin density wave (SDW) state. This SDW state is enhanced by applying a weak 
external magnetic field \cite{haug}, but still remains at short range. Meanwhile,  small Fermi pockets have been detected in quantum oscillation 
measurements carried out at high magnetic field \cite{quantum-osc,proust,sebastian} in ${\rm YBa_2Cu_3O_{6+x}}$ system (only for x$\ge$0.5, i.e, p$
\ge$0.09). The most simple explanation for these Fermi pockets is a Fermi surface reconstruction\cite{proust,Fsurf-recons}, triggered by a spin 
and/or charge density wave order breaking the TSL. The reconstruction of the Fermi surface inferred from quantum oscillation measurements could be linked to the SDW ordering reported in YBCO with and without magnetic field \cite{hinkov,haug,haug1}, but for a lower hole doping (x$<$0.5, i.e, p$<$0.09). However, one should be rather careful since the correlation length associated with the glassy SDW state are too short to account for the observed quantum oscillation and no quantum oscillation have been reported so far at the small hole doping at which the SDW is observed. 

In following sections, we will present compelling evidence of an ordered magnetic phase in the pseudogap state. We will show that this order phase displays strong similarities with the circulating current (CC) phase proposed by C.M. Varma \cite{simon,cmv06}. The CC-phase is associated with a Q=0 electronic instability and therefore preserves the TSL. Furthermore, its generalized Ising (Ashkin-Teller) phase transition does not produce any strong jump in the specific heat for a large range of parameters. While a 3D long range Q=0 magnetic order is clearly visible in the pseudogap state of system such ${\rm YBa_2Cu_3O_{6+x}}$ (YBCO) or $\rm HgBa_2CuO_{4+\delta}$ (Hg1201),  we will show that this  order is frustrated in LSCO and is likely to be in competition with a stripe instability.

\section{Evidencing circulating currents}

\subsection{Loop current order}

Beyond usual charge or spin instabilities, more exotic electronic phases, spanning various CC states \cite{cmv97,simon,cmv06,ddw}, have been proposed to account for a hidden order parameter associated with the pseudogap.  
One may have a single charge current per CuO$_2$ plaquette staggered in the neighbouring cell, referred to as a D-density wave (DDW)\cite{ddw}. The DDW state implies a doubling of the unit cell. The interest for such a state has been recently revived  \cite{Kee,DDW-Field} owing to the observation of small Fermi pockets in quantum oscillation measurements at high field. In the DDW state, charge loops give rise to orbital-like magnetic moment perpendicular to $\rm CuO_2$ planes. The AF arrangement of such orbital-like moments should be directly probed by neutron scattering diffraction. However, to date, the neutron detection a DDW state is rather scarce \cite{berlin}. 

\begin{figure}[t]
\centering
\includegraphics[width=12cm,angle=-90]{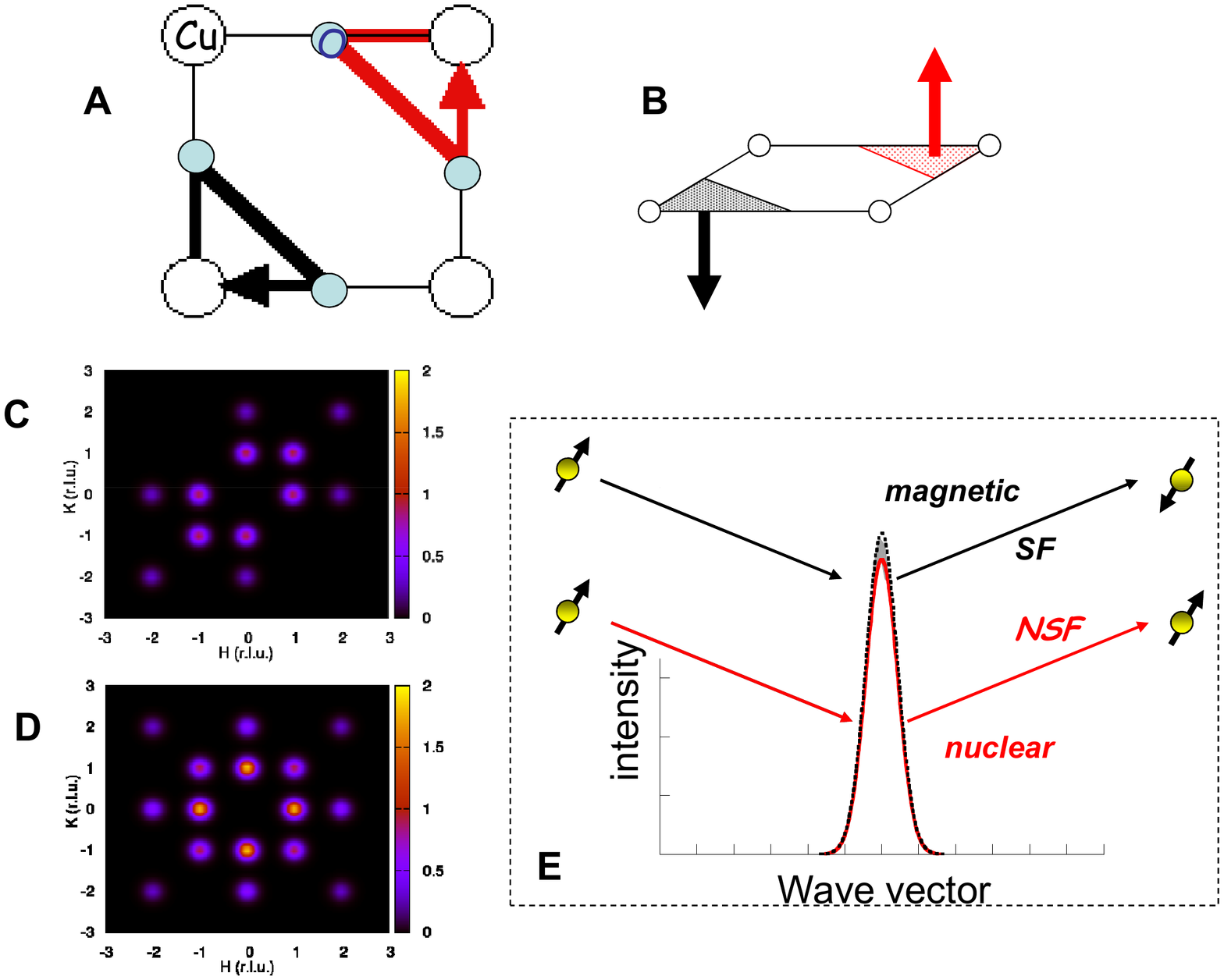}
\caption { A) Schematic description of the circulation currents flowing inside a $\rm CuO_2$ plaquettes.
In the $\rm CC-\Theta_{II}$ phase, there are two currents loops turning clockwise and anti-clockwise. Reversing the current loops and/or rotating the figure by 90$^o$ allows one to build the 4 classical domains of the $\rm CC-\Theta_{II}$ phase. B) The current loops generate staggered orbital-like magnetic moments, perpendicular to the $\rm CuO_2$ planes. C-D) Neutron scattering patterns for a C) single- and D) a multi-domain $\rm CC-\Theta_{II}$ phase, respectively. In these two figures, a typical magnetic form factor has been used to compute the magnetic structure factor. E) The $\rm CC-\Theta_{II}$ phase is a Q=0 AF phase (which belongs to the  ${\rm E_u}$ irreductible representation of the ${\rm D_{4h}}$ point group \cite{agterberg,arkady}): it breaks time reversal symmetry, but preserves lattice translation invariance. As a consequence, the nuclear and magnetic intensity are superimposed on Bragg reflections. The polarized neutron scattering technique allows one to disentangle nuclear and magnetic scatterings, which  appear selectively in the non-spin-flip (NSF) or spin-flip (SF) channels.
}
\label{Fig2}
\end{figure}

Other CC states have been predicted \cite{cmv97,simon,cmv06} from the 3-band Hubbard model involving both copper d-orbitals and in-plane oxygen p-orbitals. As a result of the Cu-O repulsion term,
a loop-current state is stabilized yielding patterns with two ($\rm CC- \Theta_{II}$ phase), or four ($\rm CC- \Theta_{I}$ phase) current loops per unit cell. The phase $\rm CC- \Theta_{II}$ corresponding to two opposite current loops is depicted in Fig.~\ref{Fig2}.A, it belongs to the  ${\rm E_u}$ irreductible representation of the ${\rm D_{4h}}$ point group \cite{agterberg,arkady}. A current flows from the Cu atom through the nearest oxygen atoms, then back to Cu. By principle, these phases break the time-reversal symmetry (TRS). The broken TRS has been indeed  observed by ARPES in $\rm Bi_2Sr_2CaCu_2O_{8+\delta}$ (Bi2212) through a spontaneous dichroism below T* \cite{kaminski}. As sketched in Fig. \ref{Fig2}.B, each closed current loop produces an orbital magnetic moment which should be measurable by neutron diffraction. A sizeable orbital magnetic moment of 0.1 $\mu_B$ is typically expected \cite{cmv97}, pointing perpendicularly to the CuO$_2$ planes. In contrast to the DDW state, these phases preserve the TSL, as the same pattern is exactly reproduced in every neighbouring cells. Because the loops are staggered within each unit cell (fig. \ref{Fig2}.B), there is no net magnetization contrary to a ferromagnetic phase. Therefore, the magnetic state corresponds to a Q=0 orbital AF order(Q=0, AFO).

%----------------------------------------------------------------------------------------------
%----------------------------------------------------------------------------------------------

Motivated by this approach, we developed a neutron diffraction experiment to evidence such a magnetic order. Fig.~\ref{Fig2}.C represents the magnetic intensity in the reciprocal space of a quadratic CuO$_2$ plane taking the magnetic structure given in Fig. \ref{Fig2}.B. As the state preserves the TSL, the magnetic scattering is only expected on top of nuclear Bragg peaks for integer values of H and K. Importantly, due to the existence of two staggered moments in each unit cell, no intensity is predicted for H=K=0. In $\rm CC-\Theta_{II}$ state, 4 classical domains can exist \cite{cmv06}. One CC state is shown in Fig.~\ref{Fig2}.A. The three other states can be obtained by reversing the current flow or by rotating Fig.~\ref{Fig2}.A by 90$^o$ \cite{cmvhe}. Two domains should produce the neutron scattering pattern given in Fig.~\ref{Fig2}.C and the two others should produce a similar scattering pattern but rotated at 90$^o$.  The scattering pattern in Fig.~\ref{Fig2}.D is obtained by assuming the same population distribution for each domain, restoring the quadratic C$_4$ symmetry in the total scattering pattern. One sees that the most intense magnetic peak would be for (H,K)=$(\pm1,0)\equiv(0,\pm1)$. 

\subsection{Polarized neutron technique}

To separate the nuclear and magnetic scatterings - superimposed at the same Bragg position - requires polarized neutron diffraction experiments. As shown in Fig.~\ref{Fig2}.E,  the nuclear scattering interaction conserves the neutron spin whereas the magnetic scattering interaction can flip the neutron spin. In a polarized neutron diffraction measurement, the incoming neutrons are in the same spin state and one analyses the spin state of the scattered neutrons. On can then  distinguish the spin-flip (SF) scattered intensity from the non-spin-flip (NSF) one. This technique is extremely powerful.
Unfortunately, the  (NSF) nuclear intensity is much larger than the (SF) magnetic intensity, typically one thousand time larger. The difficulty of the experiment then resides in the capability of producing a high polarized neutron beam, reliable and stable in time and in position. In the case where the magnetic order is short range, the magnetic intensity is redistributed in momentum space. In this limit, the technical issue is related to the weakness of the magnetic signal/background ratio, but the polarization stability is no more relevant. These experimental constraints are the best satisfied using the polarized neutron diffraction on the 4F1 triple-axis spectrometer around the reactor Orph\'ee at the Laboratoire L\'eon Brillouin (LLB), Saclay (France) \cite{fauque,mook,li1,baledent}. The observation of a magnetic scattering whatever the experimental difficulties is a significant sign of the universality  of the phenomenon. 

As discussed in  \cite{fauque,mook,li1,baledent}, the polarized neutron diffraction setup is similar to that originally described in \cite{moon} with longitudinal polarization analysis (see also refs. \cite{yvan,dhlee} in the context of high-T$_c$ cuprates). A polarized incident neutron beam is obtained with a polarizing mirror (bender), the neutron energy is fixed at $E_i=13.7$ meV. The polarization analysis is performed with an Heusler analyser. A radio-frequency Mezei flipper is put between the bender and the sample in order to reverse the incident neutron polarisation.  A pyrolythic graphite filter has been put before the bender to remove higher order neutrons. The direction of the neutron spin polarization, 
{\bf P}, at the sample position is selected by a small guide field {\bf H} of the order of 10 G. 

 In polarized neutron scattering technique, one can define the inverse of the flipping ratio (R$^{-1}$) as R$^{-1}= I_{SF}/I_{NSF}$ where $I_{SF}$ and $I_{NSF}$ stand for the SF and NSF neutron intensities, respectively. For a perfectly polarized neutron beam, R$^{-1}$ should be equal to 0 in absence of any magnetic scattering. In real experiments, this is not the case. As a results of the unperfect neutron beam polarization, there is a leakage of the NSF intensity into the SF channel. At the spectrometer 4F1, a highly polarized neutron beam is routinely obtained with R$^{-1}$ranging between 1/40 and 1/100. In absence of a magnetic order, R$^{-1}$ exhibits a smooth behaviour with temperature, ideally constant but in reality one typically observes a slight drift of R$^{-1}$ upon cooling. The origin of this temperature dependence is unclear and varies from one experiment to another. It is likely due to imperfections in the neutron polarization and the displacement of the sample in the neutron beam when changing temperature. For a better  data analysis, one needs  to care about this change of the flipping ratio with temperature. $R(T)$ is measured from the evolution of flipping ratios at Bragg peaks where the magnetic order is not present. In the presence of an additional magnetic scattering, an upturn at low temperature in R$^{-1}$  shows up. We prove that method to be efficient enough to see weak magnetic moments ($\sim 0.05 \mu_B$) on top of nuclear Bragg peaks, see e.g. the determination of the A-type antiferromagnetism in Na cobaltate systems \cite{sibel}. 

In an unpolarized neutron diffraction measurement and for a magnetism associated with unpaired electrons, the interaction between the neutron spin and the magnetic moments in the sample is of dipolar type \cite{squires}. As a consequence, 
only the components of the order moment ${\bf M}$ which are perpendicular to {\bf Q} the transferred momentum contribute to the magnetic scattering $|F_M|^2$ \cite{squires}: $|F_M|^2 \propto |{\bf M}_{\perp}|^2$ with $ {\bf M}_{\perp} = {\bf M} - ({\bf M}.{\hat{\bf Q}}) {\hat{\bf Q}}$ with ${\hat{\bf Q}}$ the unit vector along {\bf Q} ($|{\hat{\bf Q}}|$=1). In a polarized neutron diffraction, the amount of magnetic scattering in the NSF channel is proportional to $|{\bf M}_{\perp}.{\bf P}|^2$, the neutron polarization vector being normalized ($|{\bf P}|^2=1$). The rest of the magnetic scattering (i.e  $|{\bf M}_{\perp}|^2 - |{\bf M}_{\perp}. {\bf P}|^2$) appears in the SF channel. Thus, for the neutron spin polarization {\bf P}$//${\bf Q}, the full magnetic scattering appears in the SF channel.

\subsection{Samples}

Experiments were performed on about a dozen samples of either the YBCO system  \cite{fauque,mook}, or the  Hg1201 system \cite{li1}. These systems, whose crystal structures are given in Fig.~\ref{Fig6}, allow us to cover a large range of doping, especially in the underdoped regime where the pseudogap occurs.  In both systems, it is worth pointing out that the experimental results are very consistent with each other and with the expected phase diagram, whatever the origin of samples, excluding extrinsic effects caused by impurity phases as it has recently argued from a single sample \cite{sonier}. In Hg1201 system, this study has been made possible due to the recent breakthrough in single crystal synthesis of samples with mass $\sim$ 1 g \cite{zhao}.

The scattering Bragg wave vector {\bf Q}=(H,K,L) is given in units of the reciprocal lattice 
vectors, $a^* \sim b^* = 2\pi/a$  and $c^*=2\pi/c$. Most of the data have been obtained in a scattering plane where all Bragg peaks like {\bf Q}=(H,0,L) were accessible. In the YBCO, the crystal structure is orthorhombic and plain samples are  twinned such that (H,0,L)$\equiv$(0,K,L). However, by applying an uniaxial pressure at high temperature (see e.g. \cite{hinkov}), it is possible to detwin YBCO single crystals. In such a case, one can  distinguish  scattering along {\bf a}* and {\bf b}*.

%----------------------------------------------------------------------------------------------
\begin{figure}[t]
\centering
\includegraphics[width=12cm,angle=0]{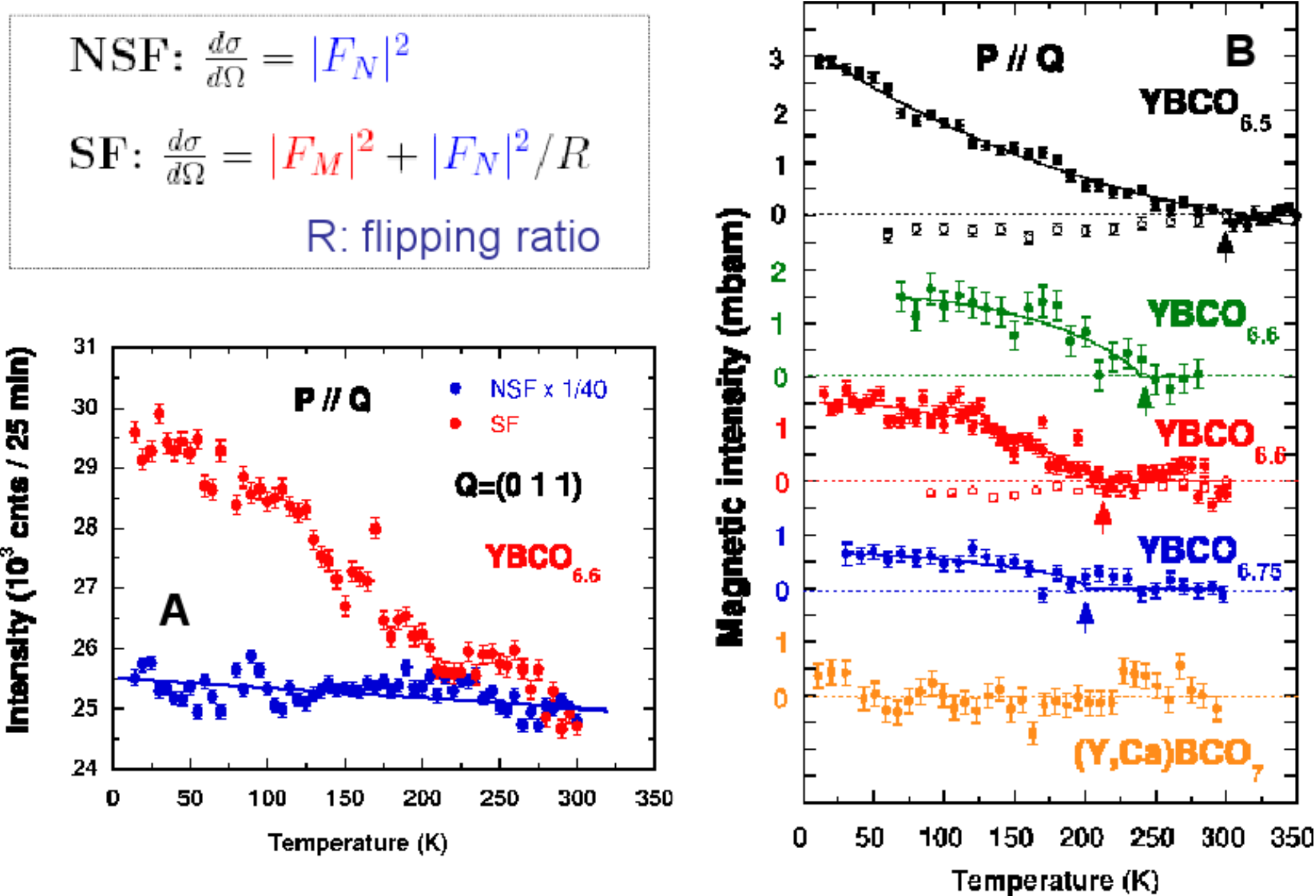}
\caption {A) $\rm YBCO_{6.6}$: Temperature dependence of the SF intensity (red)  and the NSF intensity (blue) measured at {\bf Q}=(0,1,1) for a polarization {\bf P}$//$ {\bf Q}. The formula give the expected scattering in both channels where $|F_N|^2$ and $|F_M|^2$ are the nuclear and the magnetic scattering respectively. Both curves in A were rescaled at high temperature by dividing the NSF intensity by R=40 (the flipping ratio for that experiment). B) $\rm (Y,Ca)BCO_{6+x}$: Temperature dependence of the magnetic intensity for {\bf P}$//${\bf Q} at {\bf Q}=(0,1,1) ($\equiv$ (1,0,1)) (full symbols) and {\bf Q}=(0,0,2) (open symbols). From top to bottom, the hole doping increases through the gradual change of the oxygen stoichiometry or extra substitution of Ca for Y. (adapted from \cite{fauque}).
}
\label{Fig3}
\end{figure}
%----------------------------------------------------------------------------------------------

\section{Magnetic order in the pseudogap phase}
 
\subsection{Magnetic order in YBCO and Hg1201}

As discussed above, the magnetic scattering is expected to be the largest at in-plane Bragg indices (H,K)= $(\pm1,0)\equiv(0,\pm1)$ for any integer L value  along {\bf c}*. To evidence the magnetic signal, we need to look for Bragg peaks where the structural nuclear intensity is reduced. In all cuprates, this situation is better realized for L=1. The Bragg peak intensities in both NSF and SF channels are  shown in Fig.~\ref{Fig3}.A for a detwinned YBCO$_{6.6}$ sample\cite{fauque}. The NSF intensity slightly increases with decreasing temperature as it is expected at small $\rm |Q|$ due to the Debye-Waller factor. In contrast, the SF intensity departs from this behaviour below $\rm T_{mag}$=220 K  evidencing a magnetic scattering  $|F_M|^2$ developing on top of a smooth background in the SF channel, given by the polarization leakage from the nuclear intensity $|F_N|^2/R$. From both intensities, one can estimate a ratio $|F_N|^2/|F_M|^2$ of $\sim$400 for this detwinned YBCO$_{6.6}$ sample. Typically, this ratio is larger in twinned YBCO and Hg1201 samples. Converting the relations given in Fig. \ref{Fig3}, one can deduce the magnetic scattering cross-section as it has been performed for many samples (Fig. \ref{Fig3}.B) corresponding to different doping levels ranging from an underdoped YBCO$_{6.5}$ sample to an overdoped (Y,Ca)BCO sample \cite{fauque}. These data were calibrated  in absolute units by scaling the Bragg intensity to the strong nuclear Bragg peak {\bf Q}=(0,0,4). One sees in Fig.~\ref{Fig3}.B that the magnetic scattering appears below a temperature, T$_{mag}$. Both T$_{mag}$ and  the magnitude of the magnetic scattering increase with decreasing hole doping as expected for the pseudogap phenomenon. At variance, similar analysis at {\bf Q}=(0,0,2) (empty symbols on Fig. \ref{Fig3}.B) show no additional magnetic scattering below T$_{mag}$ in agreement with the expected null  magnetic structure factor for orbital moments (Fig. \ref{Fig2}.D) related to a CC order\cite{cmv97,simon,cmv06}. At large $\rm |Q|$, the magnetic form factor is expected to considerably reduce the signal. Accordingly, measurements at the {\bf Q}=(2,0,1) reflection show no magnetic scattering \cite{mook}.

%----------------------------------------------------------------------------------------------
\begin{figure}[t]
\centering
\includegraphics[width=15cm,angle=0]{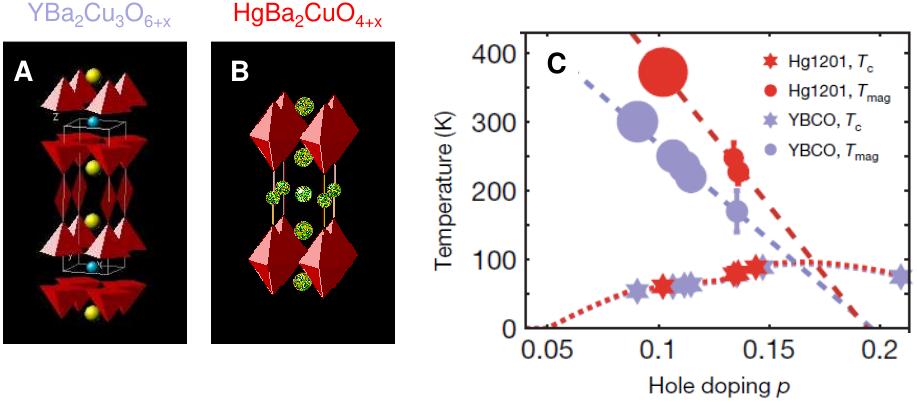}
\caption {
 A) Crystal structure of $\rm YBa_2Cu_3O_{6+x}$ (a=3.85 \AA, c=11.7 \AA). It contains 2 $\rm CuO_2$ planes per unit cell and is orthorhombic. Cu sites are embedded into oxygen pyramids and there is a dimpling of $\rm CuO_2$ planes. B) Crystal structure of $\rm HgBa_2CuO_{4+\delta}$ (a=3.87 \AA, c=9.5 \AA). There is only one $\rm CuO_2$ plane per unit cell and the system is tetragonal. Cu sites are located inside oxygen octahedra. Copper and planar oxygens lie exactly in the same plane. C) Hole doping dependence of the superconducting critical temperature $\rm T_c$ (stars) and the ordering temperature $\rm T_{mag}$  of the (Q=0, AFO) magnetic state (bullet).  Red and light blue symbols correspond to $\rm YBa_2Cu_3O_{6+x}$ (YBCO) and $\rm HgBa_2CuO_{4+\delta}$ (Hg1201) systems. The relationship between p and $\rm T_c$ has been established by a systematic study of the c lattice parameter in YBCO where the true doping can be simply estimated in the oxygen-ordered orthorhombic phases\cite{liang}. Unfortunately, the exact doping level in Hg1201 cannot be determined with accuracy. However, the $\rm T_c$ curve exhibits the same shape in doping as YBCO giving confidence that the same relationship between p and $\rm T_c$ can be used. The size of symbols is proportional to the magnitude of the ordered magnetic moment.
% The inset shows the same diagram, but using a different method\cite{liang} to estimate the hole doping (see text).
(adapted from \cite{li1}). }
\label{Fig6}
\end{figure}
%----------------------------------------------------------------------------------------------

\subsubsection{Magnetic moment}

Next, from $|F_M|^2$  (Fig. \ref{Fig3}.B), one can estimate the magnitude of the ordered magnetic moment $\rm |{\bf M}|$ defined using the magnetic neutron cross section for the $\rm CC- \Theta_{II}$ phase. Keeping in mind that there are 2 opposite orbital moments per CuO$_2$ plaquette, the neutron cross section  of the magnetic order given in Fig. \ref{Fig2}.B and for a Bragg position {\bf Q}=(H,0,L) simply reads 
%----------------------------------------------------------------------------------------
\begin{equation}
|F_M|^2 = r_0^2 f(Q)^2 \beta(L)^2  4 \sin^2(2 \pi x_0 H) | {\bf M}_{\perp}|^2 
\label{Imag}
\end{equation}
%----------------------------------------------------------------------------------------
where $r_0$=-0.54 $10^{-12}$ cm ({\it i.e.} $r_0^2=290$ mbarns) is the neutron magnetic scattering length.
$x_0=0.146$ is the position of the magnetic moment within the unit cell ({\it i.e.} the center of the triangle in Fig. \ref{Fig2}.A). 

$f(Q)$ stands for the magnetic form factor comprised by principle between 0 and 1. The calculation of the magnetic form factor necessitates in principle a detailed knowledge of the real space extension of the orbital moments, involving the Cu-d$_{x^2-y^2}$ and O-$p_{x,y}$ orbitals. So far, such a calculation has not been performed for the observed magnetic order. We take an arbitrary estimate of $f(Q)^2=0.5$ around {\bf Q}=(1,0,1). This crude assumption has been previously made for YBCO and Hg1201 \cite{fauque,mook,li1} to be able to make a first estimate of the magnetic moment. 

$\beta(L)$ is the magnetic structure factor along {\bf c}* related to the magnetic structure within the $\rm (CuO_2)_2$ bilayer (in the case of  the YBCO system). As shown from the L-dependence on three different peaks \cite{berlin,fauque,mook}, the coupling along {\bf c}* between the magnetic moments has to be ferromagnetic as the largest magnetic intensity is observed for L=0. Accordingly, a natural form for $\beta(L)$ is then $2 \cos(\pi z L)$ with $z$=0.29 is the reduced distance between CuO$_2$ layers in YBCO\cite{berlin}.

Considering, on the one hand, a magnetic moment {\bf M} with the 3 components $(M_a,M_b,M_c)$ along the main crystallographic directions and, on the other hand, the existence of 4 domains in the $\rm CC- \Theta_{II}$ phase (built from Fig. \ref{Fig2}.A),  the square of the magnetic components perpendicular to {\bf Q} reads:
%----------------------------------------------------------------------------------------
\begin{equation}
|{\bf M}_{\perp}|^2 =\frac{1}{2} [1+ (\frac{c*L}{Q})^2] M_{a,b}^2 + [1- (\frac{c*L}{Q})^2] M_{c}^2
\label{M-perp-2}
\end{equation}
%----------------------------------------------------------------------------------------
with $M_{a,b}= \sqrt{M_{a}^2+M_{b}^2}$ the magnitude of the in-plane magnetic component. At {\bf Q}=(1,0,1),  $(\frac{c*L}{Q})^2$ is equal to 0.1. If {\bf M} were perpendicular to the $\rm CuO_2$ plane, then  $|{\bf M}_{\perp}|^2$ would be  $\sim |{\bf M}|^2$. As will be shown in the next section, {\bf M} is not simply along the {\bf c} axis, but rather tilted at $\sim$ 45 $^{o}$. One gets a rough estimate $|{\bf M}_{\perp}|^2 \sim \frac{3}{4} |{\bf M}|^2$. 
 
Finally, using these various approximations, one obtains a magnetic moment of $ \sim 0.1 \mu_B$ when $|F_M|^2 \sim$ 1 mbarn in Fig. \ref{Fig3}.B. This value decreases  when increasing hole doping. It should be emphasized that this moment is given per triangular current loop in the $\rm CC - \Theta_{II}$ model of Fig. \ref{Fig2}.A and corresponds to the order of magnitude expected theoretically\cite{cmv97}. Of course, the precise value {$M$ would change depending on both the model and the accuracy of the different parameters appearing in Eq. \ref{Imag}.
However, the order of magnitude of the ordered moment is actually quite robust whenever other assumptions are made. 
In Hg1201\cite{li1}, the same expression Eq. \ref{Imag} can be used with $\beta(L)=1$ as there is a single layer per unit cell. Using similar  assumptions, the magnetic moment per triangle can reach $\sim$0.2 $\mu_B$ in the strongly underdoped state\cite{li1}.

\subsubsection{A true phase transition}

We then identify this magnetic order to the hidden order parameter of the pseudogap phase in YBCO\cite{fauque,mook}. Similar measurements in Hg1201\cite{li1} display the same doping dependence which can be summarized in Fig. \ref{Fig6}.C where the doping level is determined for both YBCO and Hg1201 using the same updated relationship between the doping and $\rm T_c$ \cite{liang}. It should be emphasized that the temperature $\rm T_{mag}$ matches the pseudogap temperature T* deduced from  resistivity measurements\cite{fauque,li1}. For both systems, the magnetic ordering temperature $\rm T_{mag}$ extrapolates  to zero at p=0.19, which matches the expected end point of the pseudogap phase as deduced from various physical properties such as magnetic susceptibility, entropy or resistivity \cite{tallon}. This points towards the existence of a quantum critical point at p=0.19. Clearly, the neutron data provides a strong support in favor of a true phase transition at T* \cite{revue,varmaQCP}. In addition, high resolution magneto-optic (Kerr effect) measurements consolidate this conclusion as it also evidences a time reversal breaking symmetry in the pseudogap phase of YBCO \cite{xia}. 

The pseudogap affects all physical properties. However, the value of T* for a given doping varies through the literature depending on the data analysis or on each experimental technique. That is actually the reason why T* has been largely considered as a crossover phenomenon. In order to compare the effect of the pseudogap on a given physical properties, one needs to know how this quantity can couple to the order parameter of the pseudogap phase. For instance, it has been argued\cite{aji-cmv} that the $\rm CC- \Theta_{II}$ theory could be mapped onto the Ashkin-Teller model (a model with a pair of Ising spins at each site which interact with neighboring spins through pair-wise and four-spin interactions). Monte Carlo simulations\cite{varmaQCP} of the Ashkin-Teller model show no sharp thermodynamic anomaly at the phase transition for a large set of parameters. That may corresponds to that observed in cuprates at T*\cite{tallon}. Furthermore, the order parameter exponent $\beta$ =0.18  \cite{mook} that one can extract from the neutron data falls into the proper range for such an Ising type model. Another example of a thermodynamic signature of the existence of a phase transition in the pseudogap state of underdoped YBCO is given by recent high-precision magnetization measurements\cite{leridon}. The temperature derivative of the uniform susceptibility indicates a  singular point at a temperature corresponding to $\rm T_{mag}$. This can be understood\cite{varmaQCP,leridon} as a biquadratic coupling of the magnetic order parameter with the uniform magnetization in a way similar to what happens in antiferromagnet.
 
%----------------------------------------------------------------------------------------------
\begin{figure}[t]
\centering
\includegraphics[width=12cm,angle=-90]{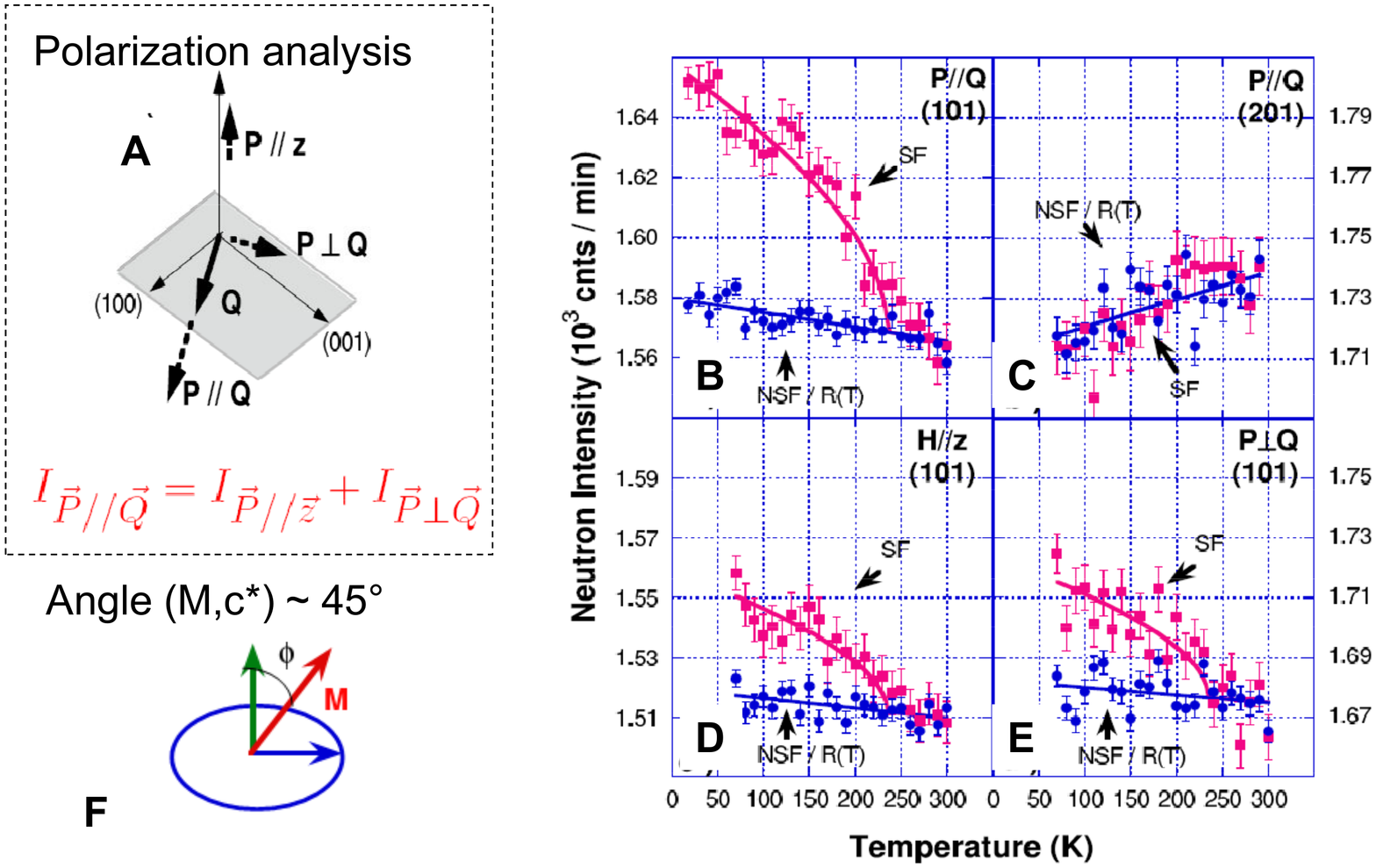}
\caption { A) Schematic description of  the three orthogonal neutron polarizations (see text). The shaded area represents the scattering plane defined by the directions (100) and (001). B-E) Temperature dependencies of the intensities in the SF channel (magenta):  B) for {\bf P} $//$ {\bf Q}  at {\bf Q} =(1,0,1)  and (C) at {\bf Q}=(2,0,1), (D) for {\bf P} $//$ {\bf z}   and (E) {\bf P} $\perp$ {\bf Q} at {\bf Q}=(1,0,1). For all panels, blue symbols indicate the polarization leakage, given by the NSF intensity divided by the temperature dependence of the flipping ratio, R(T). R(T) is found by a fit in panel C of the ratio NSF/SF of the Bragg peak {\bf Q}=(2,0,1) as R(T)=R(300 K)\{1+0.02(1-T/300) F) Orientation of the magnetic moment with respect to the {\bf c} axis (green arrow), as deduced from the polarization analysis (adapted from \cite{mook}).
}
\label{Fig5}
\end{figure}
%----------------------------------------------------------------------------------------------

\subsubsection{Polarization analysis}

Polarized neutron  diffraction technique allows us  to determine the direction of the magnetic moment through a polarization analysis\cite{moon}.  As the interaction of the neutron spin with magnetic moments is of dipolar type, only the magnetic components perpendicular to the wavevector {\bf Q} are measurable\cite{squires}. Furthermore, only a magnetic component perpendicular to the neutron spin polarization {\bf P} is spin-flip whereas a component along {\bf P} is non-spin-flip\cite{squires}.  The amount of magnetic scattering in the SF channel therefore depends on the choice of {\bf P}.

For a polarized neutron analysis, one usually introduces a set of three orthogonal directions:  The $\bf\hat{x}$ axis is parallel to {\bf Q}, the $\bf\hat{y}$  axis is perpendicular to {\bf Q} in the scattering plane and the $\bf\hat{z}$  axis is perpendicular to the scattering plane.  For a neutron spin polarization along $\bf\hat{x}$ ({\bf P}$//${\bf Q} $\equiv$ {\bf H}$_x$), the full magnetic intensity appears in the SF channel: $I_{\bf{P}// \bf{Q}} \propto M_{\perp y}^2+ M_ {\perp z}^2$, $M_{\perp y}$ and $M_{\perp z}$ are the components of ${\bf M_{\perp}}$ along $\bf\hat{y}$ and $\bf\hat{z}$  respectively. For a neutron spin polarization  applied along $\bf\hat{y}$ ({\bf P}$\perp${\bf Q} $\equiv$ {\bf H}$_y$), only one part of the magnetic intensity appears in the SF channel: $I_{\bf{P} \perp \bf{Q}} \propto M_{\perp z}^2$.  The complementary magnetic intensity $I_{\bf{P}// \bf{z}} \propto M_{\perp y}^2$ shows up in the SF channel when the neutron spin polarization is applied along $\bf\hat{z}$ ({\bf P}$//${\bf z} $\equiv$ {\bf H}$_z$). As a result, in the SF channel, one can deduce a specific polarization sum rule for the neutron intensity: $I_{\bf{P} / /\bf{Q}} = I_{\bf{P}\perp \bf{Q}} + I_{\bf{P}//\bf{z}}$. It is worth noticing that this polarization sum rule is {\it only} fulfilled for a magnetic scattering in the absence of chirality in the system. 

 %----------------------------------------------------------------------------------------------
\begin{figure}[t]
\centering
\includegraphics[width=12cm,angle=0]{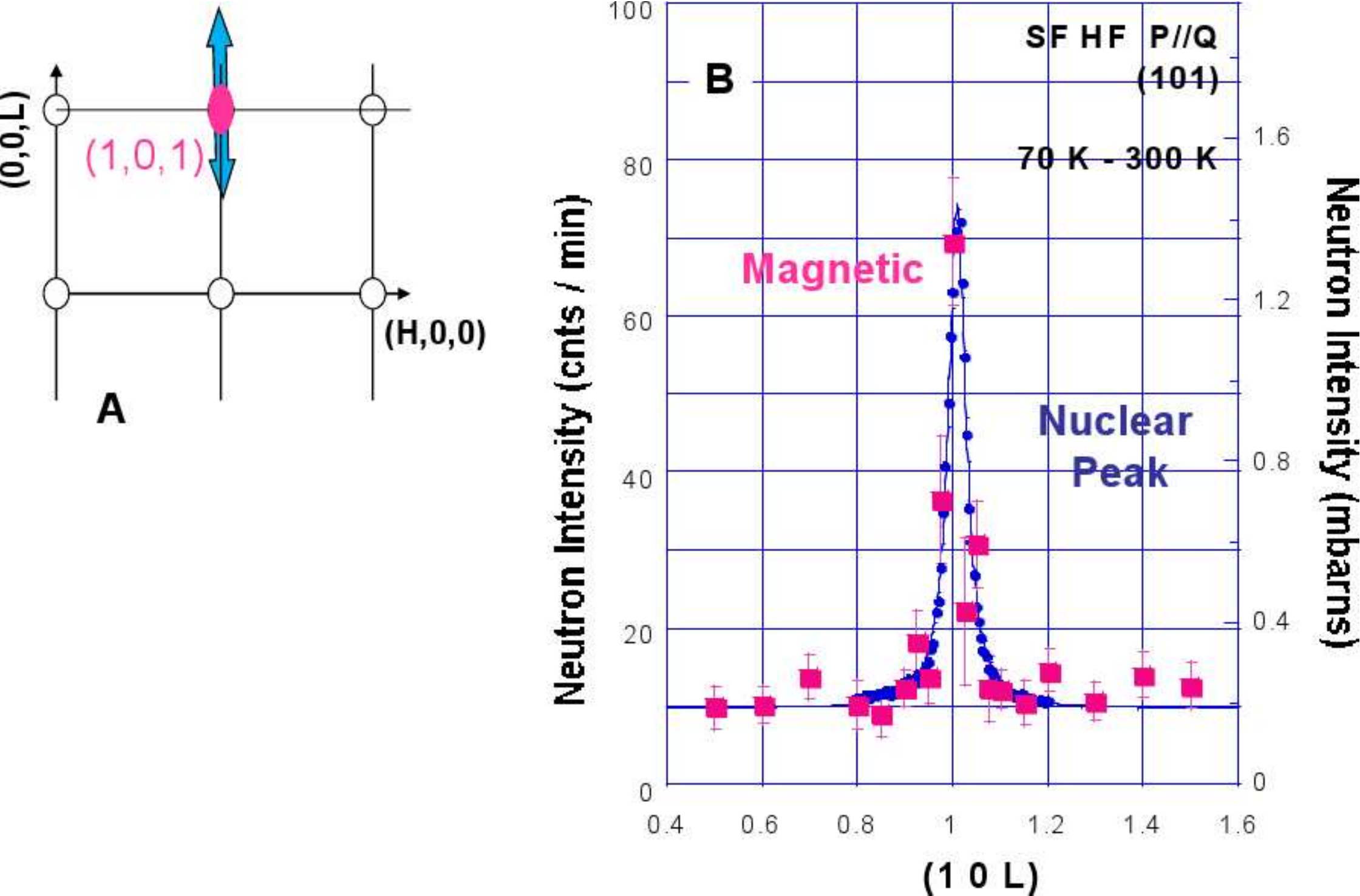}
\caption { $\rm YBCO_{6.6}$: A) [100]/[001] scattering plane and schematic description of a scan along the [001] direction across {\bf Q}=(1,0,1). B) Difference between L-scans at 70 K and 300 K in the SF channel and for polarisation {\bf P}$//${\bf Q} (magenta). The magnetic signal, centered at {\bf Q}=(1,0,1), is resolution limited. The blue symbols stand for the resolution limited nuclear Bragg scattering at {\bf Q }=(1,0,1) in the NSF channel (adapted from \cite{mook}).
}
\label{Fig4}
\end{figure}
%----------------------------------------------------------------------------------------------

The polarization analysis has been performed in YBCO\cite{fauque,mook} for the Bragg peak {\bf Q}=(1,0,1) in twinned samples as shown in Fig. \ref{Fig5}. In first approximation, the flipping ratio is constant in temperature\cite{fauque,li1}. To improve the data analysis, one needs further to fit it empirically by a straight line \cite{mook,li3} as it has been done in Fig. \ref{Fig5}.C. First, the magnetic scattering is larger for {\bf P}$//${\bf Q} as it should be. Next, the polarization sum-rule above is fulfilled proving the magnetic nature of the signal observed below $\rm T_{mag}$.  The same conclusion has been also demonstrated in another YBCO$_{6.6}$ sample \cite{fauque} as well in recent data obtained in Hg1201 \cite{li3}. 

The data reported in Fig.~\ref{Fig5} were measured in the scattering plane given by the directions (100) and (001): the $\bf\hat{z}$ axis therefore corresponds to the (010) direction. The orbital magnetic moment in a CC-phase is then expected to be parallel to {\bf c} (Fig.~\ref{Fig2}.B). Thus, one should expect $I_{\bf{P} / /\bf{Q}}= I_{\bf{P}//\bf{z}}$ and $ I_{\bf{P}\perp \bf{Q}}$=0. At variance,  one observes a magnetic contribution for all neutron polarizations (Fig.~\ref{Fig5}.B,D,E). More specifically, $I_{\bf{P} / /\bf{Q}}$=1.4 mbarn and  $I_{\bf{P}\perp \bf{Q}}=I_{\bf{P}//\bf{z}}$=0.7 mbarn in the YBCO$_{6.6}$ sample\cite{mook} shown in  Fig.~\ref{Fig5}. Qualitatively, the fact we observe a non-zero magnetic intensity for {\bf P}$//${\bf z} (Fig.~\ref{Fig5}.D) proves that a large part of the magnetic moment is pointing along {\bf c}. However, the intensity for  {\bf P}$//${\bf z} is smaller than the one for {\bf P}$//${\bf Q} indicating that an in-plane component is also present. 

Next, to discuss the magnetic moment direction, we use the same simple model used in previous sections to evaluate the order of magnitude of the ordered magnetic moment. We assume a model with collinear moments pointing along a generic direction given by, {\bf M}=$(M_a,M_b,M_c)$  with no preferential in-plane direction (Fig. \ref{Fig5}.F). One then needs to  calculate the neutron intensity from Eq.~\ref{Imag} and Eq.~\ref{M-perp-2} for {\bf Q}=(1,0,1) and for each neutron polarization \cite{squires}. That gives $I_{{\bf{P}\perp \bf{Q}}} \propto [ \frac{1}{2} M_{a,b}^2 ]$, $I_{{\bf{P}//\bf{z}}} \propto [ 0.05 M_{a,b}^2 + 0.9 M_c^2 ]$ and the sum of both terms for ${\bf{P}//\bf{Q}}$. One can define the angle, $\phi$ between the magnetic moment and {\bf c} (see Fig. \ref{Fig5}.F) which can be written as $\tan(\phi)= \sqrt{M_{a,b}^2/M_c^2}$. Using the result of Fig. \ref{Fig5} at {\bf Q}=(1,0,1), one gets $\phi = 55 \pm 7 ^{o}$  \cite{mook}. At {\bf Q}=(1,0,0), the angle is found to be smaller 
 $\phi = 35 \pm 7 ^{o}$ \cite{mook}.
 It is unclear at the moment what is the meaning of such a difference. The given errors are related to statistical error, however, due to possible systematic bias in the data analysis, the actual error might be larger. One can then give a conservative estimate of  $\phi = 45 \pm 20 ^{o}$ valid over few samples and Bragg spots \cite{fauque}. The recent data in Hg1201 leads to the same tilt of the moment relative to the {\bf c} axis \cite{li3}. One then sees that this effect does not depend on the number of CuO$_2$ per unit cell but is intrinsic to that novel (Q=0, AFO) magnetic order. The implication of this finding will be discussed in the last section. 

To estimate the angle  $\phi$, we here assumed a collinear magnetic structure for simplicity. However, it should be noticed that more complex magnetic structures might exist, in particular those involving non-collinear moments. In such cases, different magnetic diffraction patterns could be obtained for each magnetic component in the neutron scattering cross-section. In principle, this might yield a different estimate of the tilted angle. 

\subsubsection{Long range order}

So far, only the intensity at the Bragg position has been discussed. One crucial question about this (Q=0, AFO) broken symmetry is whether the order is at long range or not. To answer that question, one needs to perform scans in the reciprocal space without loosing the polarization condition. At the level of accuracy necessary to seize the magnetic peak, this is actually only 
working properly along {\bf c}* (sketched by the blue arrow in Fig.~\ref{Fig4}.A) which almost corresponds to a rocking scan of the sample. Fig.~\ref{Fig4}.B depicts the magnetic scattering, obtained by taking a temperature difference of the SF scattering for the  {\bf P}$//${\bf Q}. The blue points represent the non-magnetic NSF Bragg scattering scaled to the same amplitude as the magnetic scattering. The q-width of the NSF Bragg peak  is given by the resolution of the spectrometer: here, the full width at half maximum is 0.013 \AA$^{-1}$. The SF and NSF intensities superpose very well. This implies that the q-width of  magnetic peak is also limited by the spectrometer resolution, suggesting that the magnetic state is ordered at long range along {\bf c} \cite{fauque,mook}. From the resolution width, one can determine a lower bound for the magnetic correlation length of 75 \AA.

%----------------------------------------------------------------------------------------------
\begin{figure}[t]
\centering
\includegraphics[width=15cm,angle=0]{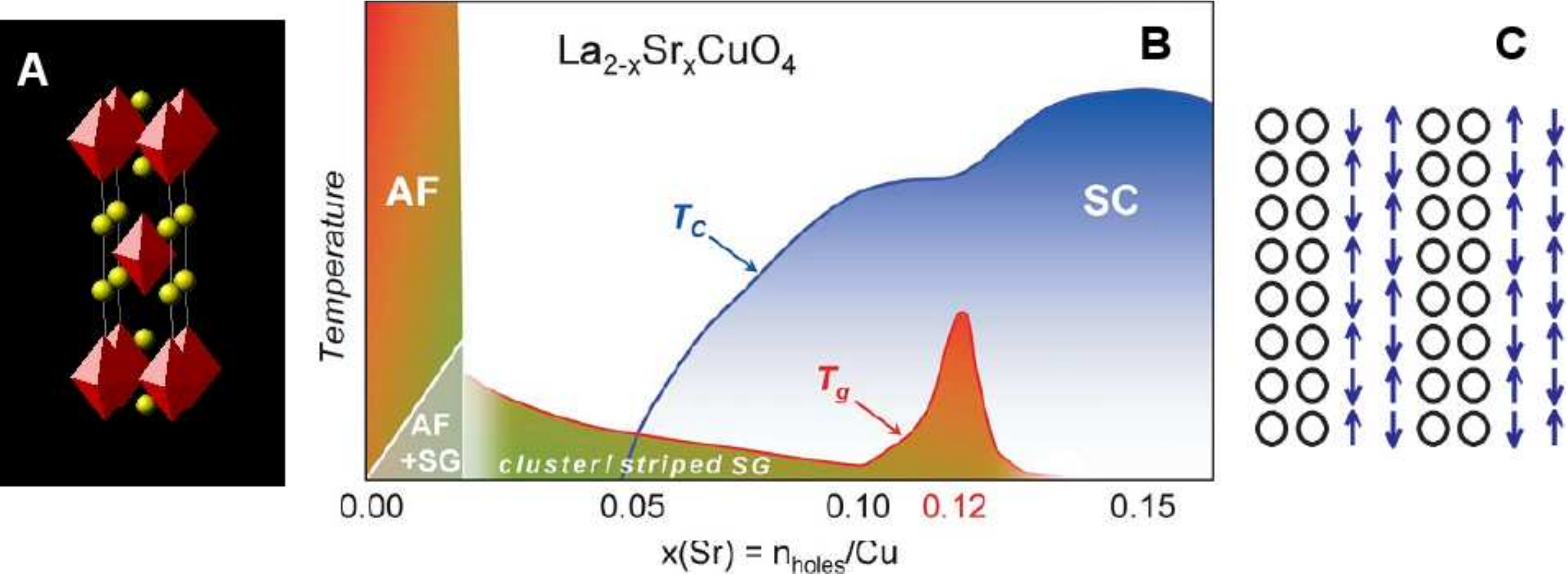}
\caption {A) Body centered crystal structure of monolayer system $\rm La_{2-x}Sr_xCuO_4$. The hole doping is achieved by substitution of Sr$^{2+}$ for La$^{3+}$ and the hole doping (p) is equal to the Sr content (x). In the low temperature orthorhombic (LTO) phase, the orthorhombicity appears along the [110] direction in tetragonal notation. B) Phase diagram of $\rm La_{2-x}Sr_xCuO_4$ system as a function of hole doping. In the low doping regime below p=0.05, one observes 2 states: the antiferromagnetic (AF) state and the cluster glass state with diagonal stripes. In the metallic regime above p=0.05, the system becomes superconducting below $\rm T_c$ and below the freezing temperature $\rm T_g$ a cluster glass state with longitudinal stripes can be observed. This phase becomes more stable upon approaching the critical hole doping p=1/8 and always tends to develop at the expense of the superconducting state (from \cite{mitrovic}) C) Cartoon picture showing a bond centered longitudinal stripe state close to p=1/8. 
}
\label{Fig8}
\end{figure}
%----------------------------------------------------------------------------------------------

\subsection{The case of ${\rm La_{2-x}Sr_xCuO_4}$}

The monolayer system ${\rm La_{2-x}Sr_xCuO_4}$ (LSCO) is the archetypal high-temperature copper oxides superconductors, mostly because by changing continuously the Sr content  one can cover all the regimes of the cuprates phase diagram from the insulating AF state upto the overdoped state with Fermi-liquid properties. However, the maximum SC temperature reaches at most 37 K, a much smaller temperature than in other monolayer system such as Hg1201 (97 K). In comparison with systems like YBCO or Hg1201, LSCO displays many drawbacks. As a function of temperature, the system undergoes a structural transition from a high temperature tetragonal phase (HTT) to a low temperature orthorhombic phase (LTO). The hole doping is achieved by cationic La$^{+3}$/Sr$^{+2}$ substitution. Such a substitution reduces the tilt of the oxygen octahedra which is at the origin of the structural distortion.  The impact of this structural transition in transport or thermodynamic measurements often blurs an accurate determination of properties like the pseudogap temperature.
Furthermore, the Sr$^{+2}$ ions, located close to $\rm CuO_2$ planes, can introduce random electrostatic potentials within the planes. LSCO is therefore a rather disordered system: this can be seen in the resistivity measurements, characterized by a large residual resistivity \cite{Rullier} and in  NMR data, where the NMR linewidth are much larger than in cuprate whose $\rm T_c$ is about 90 K at optimal doping \cite{bobroff2}. In many cuprates families,  an upturn of the resistivity  can be observed at low hole doping on cooling down \cite{Ando-PRL95} . Such an upturn is characterized by a change of sign of the first derivative of resistivity as a function of temperature ($\frac{\d \rho}{dt} < 0$), as observed in semiconductors or insulators. In addition, when applying a large external magnetic field in order to reduce or remove superconductivity, this insulating-like behavior at low temperature can be detected up to a critical hole doping p$_m$, above which a metallic behavior is recovered \cite{Boebinger-PRL96}. p$_m$  is not a universal critical hole doping among cuprates: in LSCO, p$_m$ is as large as 0.16, corresponding to the optimal doping. In less disordered materials, such as YBCO, p$_m$  remains smaller than 0.09 \cite{sebastian}. The peculiarities of LSCO system appear also in characterization of the pseudogap state. While in YBCO, T* can be easily determined from resistivity measurements or NMR knight shift,  the characteristics of the  pseudogap in LSCO are typically ill-defined \cite{timusk}.

Beyond, its intrinsic disorder,  LSCO system is characterized by a pretty strong tendency towards the formation of inhomogeneous spin and charge distributions in the $\rm CuO_2$ planes, usually associated with 1D stripes \cite{stripes}.
Stripes can be viewed as filamentary charge organisations, separating AF domains in anti-phase. A picture of bond centered stripes is given in Fig.~\ref{Fig8}.C for a doping of 1/8, doped holes are confined into 2-leg ladders, separating hole poor domains with an antiferromagnetic arrangement of S=1/2 Cu spins. The 2-leg ladders with doped holes behave as antiphase boundaries between hole poor AF spin stripes. 
In LSCO, the low energy spin excitation spectrum is dominated by strong incommensurate (IC) spin fluctuations around the planar antiferromagnetic (AF) wavevector at $\rm Q_{IC}= Q_{AF} \pm (\delta,0) \equiv Q_{AF}\pm (0,\delta)$ \cite{yamada} with $\rm Q_{AF}= (0.5,0.5)$. These low energy IC fluctuations  are usually interpreted as the fingerprint of fluctuating stripes. Local probes (NMR, $\mu$SR,)\cite{julien,mitrovic} have established that these fluctuations can become static at very low temperature. These observations suggest that fluctuating stripes can  either freeze at low doping or can also be pinned down by disorder, yielding a glassy static stripe phase. As shown in Fig. \ref{Fig8}.B, this tendency towards stabilization of static stripes is reinforced around a doping of 1/8 where stripes are better pinned to the lattice. In isostructural compounds ${\rm La_{2-x}Ba_xCuO_4}$ \cite{fujita} and ${\rm (La,Nd)_{2-x}(Sr,Ba)_xCuO_4}$ \cite{Tranquada-review}, where superconductivity is strongly reduced, the situation is even clearer as a spin density wave (SDW) order develops at $\rm Q_{IC}$ around 30 K, in addition to a charge density wave (CDW) at 2$\rm Q_{IC}$ at higher temperature \cite{Tranquada-review}. 

LSCO is then a very interesting and complex system at the crossroad of many different electronic instabilities. What might be the  Q=0 AF0 phase in that system has to be answered. Therefore, we looked recently for the Q=0 magnetic order in a LSCO sample with x=0.085 with $\rm T_c$=22 K \cite{baledent}.

%----------------------------------------------------------------------------------------------
\begin{figure}[t]
\centering
\includegraphics[width=10cm,angle=0]{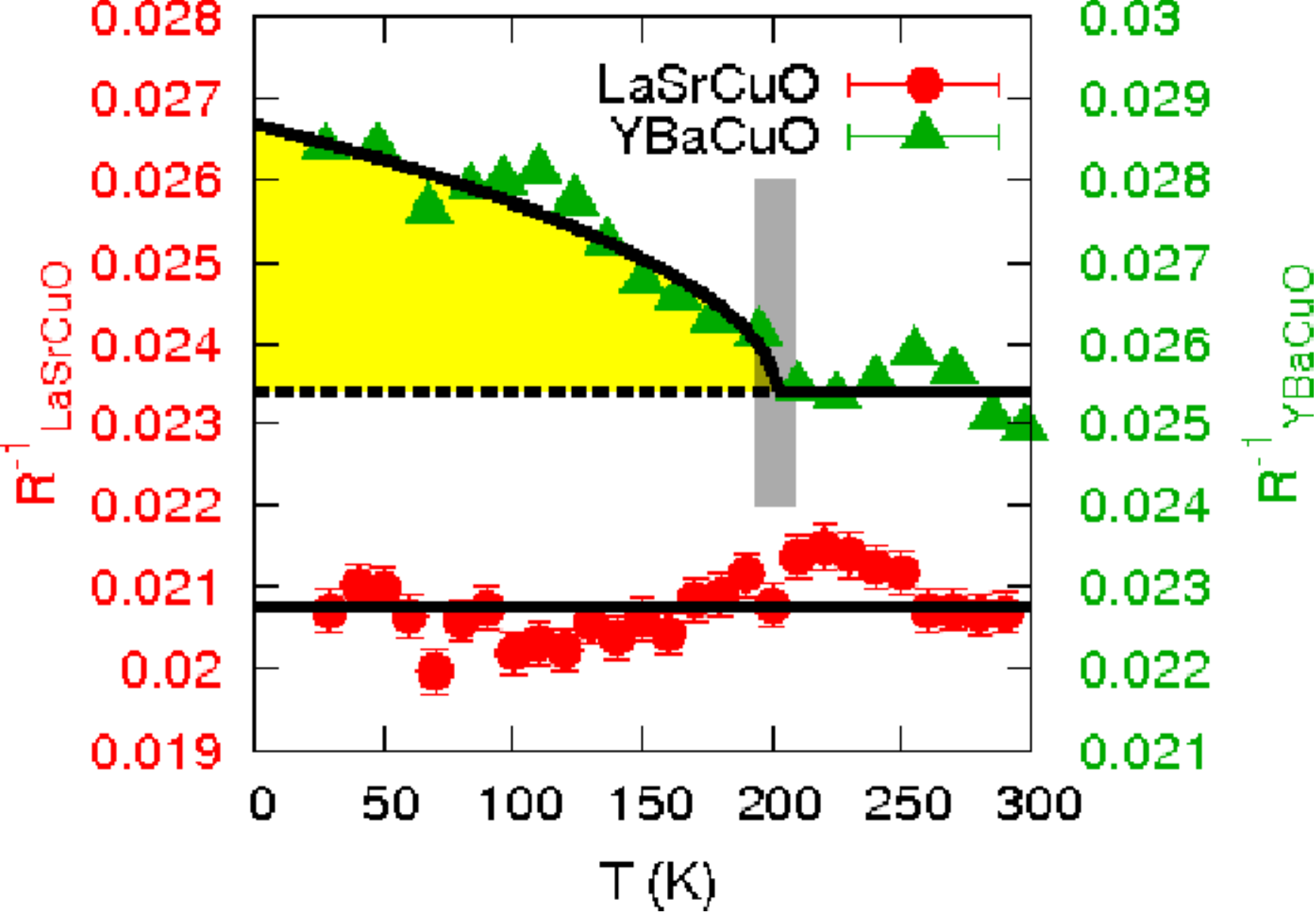}
\caption {
 Temperature dependencies of the inverse of the flipping ration ($R^{-1}$) measured at {\bf Q}=(1,0,1) for {\bf P} $//$ {\bf Q} : for $\rm YBa_2Cu_3O_{6.6}$ (YBaCuO - p=0.12 - green symbols from \cite{fauque}) and $\rm La_{2-x}Sr_xCuO_4$ LaSrCuO - x=0.085 - red symbols). The upturn of $R^{-1}$ below $T_ {mag}$=220 K indicates the appearance of the 3D Q=0 AF order upon cooling down. This feature is absent in T-dependence for $\rm La_{2-x}Sr_xCuO_4$ system, pointing out the absence of the 3D Q=0 AF order.
}
\label{Fig9}
\end{figure}
%----------------------------------------------------------------------------------------------

\subsubsection{Absence of long range order in LSCO}

Similar measurements as YBCO and Hg1201 have been repeated in LSCO. The magnetic moments of the Q=0 AFO always scatter neutrons at the same positions as the Bragg peaks of the crystallographic structure, but in contrast to YBCO and Hg1201, the crystal structure of LSCO is not primitive with a stacking along {\bf c} with CuO$_6$ octahedra shifted along the lattice diagonal (Fig.~\ref{Fig8}.A). Resulting from the body centered tetragonal phase at high temperature, the low temperature structural phase is  face centered orthorhombic: Bragg positions like (1,0,L) with even L are forbidden. The search of the long range Q=0 magnetic has therefore been carried at the Bragg point {\bf Q}=(1,0,1). As in other cuprates \cite{fauque,mook,li1}, this Bragg position offers the best compromise for a weaker nuclear Bragg peak having the proper symmetry for the Q=0 AFO. A long range 3D Q=0 magnetic order would give rise to an upturn of the inverse flipping ratio ($R^{-1}$) on cooling down.  In Fig.~\ref{Fig9}, we report on the same scale $R^{-1}$ for LSCO and $\rm YBCO_{6.6}$ \cite{fauque} for which a magnetic signal is clearly visible below 220 K.
The temperature dependence of $R^{-1}$ for LSCO only shows some fluctuations due to minute changes of the polarization of the spectrometer. Therefore, the orbital-like long range order present in other cuprates is either absent in LSCO or too weak to be experimentally detected. As nuclear Bragg intensities  in LSCO and YBCO are similar and using the error bars of Fig.~\ref{Fig9}, one can give an upper estimate of less than 0.02 $\mu_B$ for a 3D ordered orbital-like moment in LSCO, instead of $\sim$ 0.1 $\mu_B$ in YBCO$_{6.6}$ for a hole doping p$\sim$ 0.12 \cite{fauque}. 

%----------------------------------------------------------------------------------------------
\begin{figure}[t]
\centering
\includegraphics[width=12cm,angle=-90]{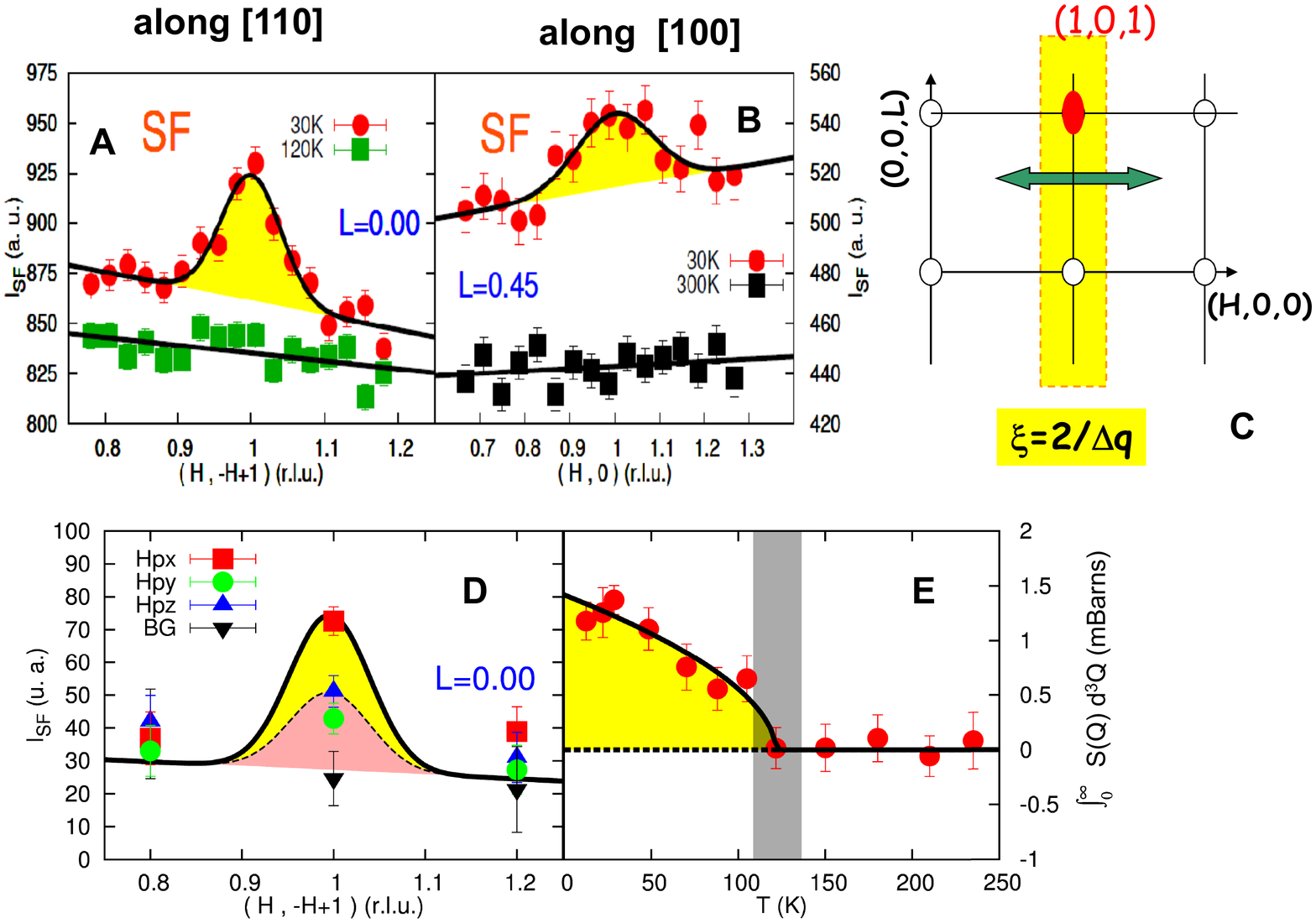}
\caption { Evidence for a short range 2D Q=0 AF order in $\rm La_{2-x}Sr_xCuO_4$: A) Diagonal scans across {\bf Q}=(1,0,0) in the SF channel for {\bf P}$//${\bf Q}. B) Longitudinal scans across {\bf Q}=(1,0,0.45) in the SF channel for {\bf P}$//${\bf Q}. C) Schematic description of a longitudinal scan across the magnetic rod (1,0,L) in the [100]/[001] scattering plane. At Low temperature, scans show the appearance of a signal centered at {\bf Q}=(1,0,L) on top of a smooth background. The signal exhibits a Gaussian profile, whose intrinsic full width at half maximum ($\Delta q$) allows an estimation of the magnetic correlation length $\xi$ (=$ 2/ \Delta q$). D) Polarized neutron scattering analysis showing the magnetic intensity in the SF channel for {\bf P}$//${\bf Q} (Hpx - red squares), {\bf P}$\perp${\bf Q} (Hpy - green bullets) and {\bf P}$//${\bf z} (Hpz - blue triangles). The black symbols stand for the SF background (BG), which does not depend on the neutron spin polarization. E/ Temperature dependence of the magnetic intensity, calibrated in absolute units (mbarns) (adapted from \cite{baledent}).}
%The solid line correspond to its fit with the function $I=I_0 (1-T/T_{mag})^{2 \beta}$ with $\beta=0.3$.} 
\label{Fig10}
\end{figure}
%----------------------------------------------------------------------------------------------

\subsubsection{Short range and 2D orbital-like order}

However, there are other possibilities for Q=0 AFO as shown in Fig.~\ref{Fig10}.C: The magnetic scattering can be either present at forbidden Bragg points for the structure, or even exist at any position along  {\bf c}*. We checked experimentally these two scenarios  \cite{baledent}.  We first studied the magnetic scattering around  the forbidden Bragg reflection (1,0,0). Fig.~\ref{Fig10}.A shows a scan in the spin-flip channel for {\bf P}$//${\bf Q} along the diagonal [1,-1,0] direction at 30 K and 120 K. A broad peak is observed there although no peak is seen in the NSF channel \cite{baledent}. The observation of such a magnetic intensity at a forbidden Bragg peak may mean that the stacking of the orbital-like moments does not preserve the lattice symmetry. The low temperature scan can be fitted by a Gaussian peak on top of a sloppy background. From the deduced peak width $\Delta_q= 0.11 \pm 0.02$ r.l.u , one obtains a short correlation length of
$\xi_{110} \equiv 2/\Delta_q \approx 11 \pm 2$\AA\ after resolution deconvolution. The magnetic scattering  does not appear  at the forbidden Bragg positions only. Indeed, additional scans at non-integer L exhibit a similar broad peak independently of L \cite{baledent}. For instance, we report on Fig.~\ref{Fig10}.B the scan for L=0.45 along the [1,0,0] direction in which the signal appears to be broader along that direction yielding $\xi_{100}= 8 \pm 3$ \AA. It should emphasized that a similar signal is also expected at the Bragg position L=1 but the signal is too small to be detected even with polarized neutron diffraction. In contrast to YBCO and Hg1201, the magnetic signal in LSCO is then  at short range and bidimensional (2D) as sketched by the yellow area in Fig.~\ref{Fig10}.C. but it is still localized around the same planar wavevectors respecting the translation symmetry of the $\rm CuO_2$ plane, say {\bf Q}$_{2D}$=(1,0), corresponding to the Q=0 AFO. 

Fig.~\ref{Fig10}.D depicts the polarization analysis of the short range magnetic order in LSCO. As in YBCO and Hg1201, the magnetic scattering occurs for all neutron polarization directions but is weaker than for {\bf P}$//${\bf Q}, fulfilling the above-mentioned sum rule. The fact that $I_{\bf{P}\perp \bf{Q}} \simeq I_{\bf{P}//\bf{z}}$ means that the moments are pointing in any preferential direction. This can be explained by considering either a rather disordered magnetic state or a similar situation as in the case of YBCO \cite{fauque,mook} and Hg1201 \cite{li1,li3}, where the moments are tilted from the {\bf c}* axis by $\approx$ 45$^{o}$. 
  
The reported 2D magnetic intensity  is rather weak with a ratio signal/background of about only 5\% \cite{baledent}. One then can ask whether this magnetic intensity is sufficient to produce a large enough local moment.  For a 3D long range magnetic order, the magnetic intensity at a magnetic Bragg position is directly proportional to square of the magnetic moment (Eq.~\ref{Imag}). For a 2D short magnetic signal, the magnetic intensity is redistributed in momentum space. In that case, this is the integrated magnetic intensity over the momentum space that scales with the square of the local magnetic moment. At first, one needs to put the 2D magnetic intensity, $I_{mag}$, in absolute units after taking into account the spectrometer resolution. $I_{mag}$ is described as 
%-----------------------------------------
\begin{equation}
I_{mag}(Q)=I_0\exp \Big[ -\ln(2) \xi^2 |{\bf Q}-{\bf Q}_{2D}|^2 \Big]
\label{sq}
\end{equation}
%-----------------------------------------
where $\xi$ is the magnetic correlation length assumed to be isotropic within the (H,K) plane for simplicity. {\bf Q}$_{2D}$=(1,0) is the planar wavevector around which the magnetic scattering occurs. $I_{mag}$ is L-independent and assumed to be static at the neutron energy scale, meaning with a characteristic energy lower than $\sim$ 1 meV. The 2D magnetic intensity, $I_{mag}$, is then convolved by the Gaussian resolution of the spectrometer. Its amplitude is further normalized in absolute units by comparison with the nuclear Bragg peak {\bf Q}=(1,0,1). Then the  Q-integrated magnetic intensity, $S_{mag}$ is obtained by integration of Eq.~\ref{sq} as $S_{mag}=\int d^3Q I_{mag}(Q)/\int d^3Q= (a/\xi)^2 I_0/(4\pi \ln(2))$  where a=3.82 \AA\ is the planar lattice parameter. Taking a mean value of $\xi$=10 \AA\ and assuming that it is temperature independent, one obtains the integrated magnetic intensity, $S_{mag}$.  As shown in Fig.~\ref{Fig10}.E,  $S_{mag}$ saturates at 1.2 mbarns at the lowest temperature, a value equivalent to the Bragg magnetic peak intensity in YBCO (Fig. \ref{Fig3}.B) \cite{fauque,mook}. 

Next, one can determine the local magnetic  moment $M_{loc}$ following the magnetic neutron cross section (Eq.~\ref{Imag}) applied to the case of LSCO meaning i) $ | {\bf M}_{\perp}|^2  =2/3  M_{loc}^2$ corresponding to a distribution of static magnetic clusters without any preferential orientation of magnetic moments from one magnetic domain to another , and ii) $\beta (L)$=1 because LSCO has a single CuO$_2$ per unit cell. Using the same mere assumptions as for YBCO and Hg1201, the magnetic local moment $M_{loc} \sim 0.1 \mu_B$ is obtained, remarkably, of similar amplitude as the long  range order {\bf Q}=0 AFO in both YBCO \cite{fauque,mook} and Hg1201 \cite{li1}.

The temperature dependence of this integrated magnetic intensity, displayed in Fig.~\ref{Fig10}.E \cite{baledent}, shows a transition $\rm T_{mag}$  around 120K, in agreement with the absence of a signal in high temperature scans.  In YBCO and Hg1201, $\rm T_{mag}$  is associated with the pseudogap temperature $\rm  T^{\star}$ \cite{fauque,li1}. According to the usual phase diagram of cuprates (Fig.~\ref{Fig1}), a much larger value for $\rm  T^{\star}$ is in principle expected. However, one the one hand, the pseudogap properties in LSCO is less accurate than in other cuprates\cite{timusk}: that could be related to the observed 2D state, characterized by finite correlation length. On the other hand, the sample doping is quite low (8.5~\%): the determination of the pseudogap behaviour is hampered by  the proximity of the metal-insulating crossover as well as by the existence of strong AF fluctuations\cite{yamada}. Actually, several anomalies have been reported close to $\rm T_{mag}$ for LSCO samples in the same doping range: these anomalies appear in the specific heat  \cite{oda,ido}, the uniform spin susceptibility \cite{oda} and the Nernst effect \cite{ong}. However, the exact relation between these various anomalies with the short range magnetic order is not settled yet. 

\section{Discussion}

In three different cuprates families, we observe a novel magnetic order in the pseudogap state. Its order parameter breaks time reversal symmetry in agreement with the observation of spontaneous Kerr effect in YBCO \cite{xia} and of dichroism in the ARPES spectra in another cuprate, Bi2212 \cite{kaminski}. Likewise this novel magnetic order preserves the translation symmetry of the lattice, being then categorized as a Q=0 antiferromagnetic order. What are the implications of this symmetry ? Independently of any theory, it proves that at least two antiparallel magnetic moments per unit cell are necessary to model our observations. Cu spins, which are usually invoked to explain magnetism in cuprates, cannot account for the novel magnetic phase as there is only Cu spin per unit cell. Its magnetic ordering gives rise to additional Bragg peaks 
(typically incommensurate) near the (1/2,1/2) wave vector\cite{Tranquada-review,hinkov,haug1}. Our finding implies that one needs to reconsider the role of oxygen orbitals, especially the in-plane p-orbitals. Theoretically, that questions the standard effective one-band Hubbard model generally considered to encompass the physics of high-$T_c$ cuprates. A priori, our observation implies that a proper starting point has to be at minimum a three-bands Hubbard model. 
%However, it has been argued that staggered orbital currents preserving translation symmetry can be also obtained in one band Hubbard model  \cite{stanescu}. 

\subsection{Orbital or spin moments ?} 

The observation of a Q=0 order in the pseudogap state of cuprates has fundamental implications. In order to account of such novel magnetic state, a decoration of the unit cell is necessary with at least two opposite magnetic moments within each squared $\rm CuO_2$ plaquette. Indeed, together with the absence of magnetic moments in magnetization measurements \cite{leridon}, the zero intensity for H=K=0 ruled out ferromagnetism as the origin of the magnetic structure. Our studies therefore suggest that a single $\rm CuO_2$ plaquette contains  internal degrees of freedom, which have been ignored in most of theoretical models developed for cuprates. This viewpoint has been recently experimentally comforted by the observation of electronic nematicity in Bi2212 by STM \cite{lawler} which also suggests intra unit cell electronic patterns. 

Of course, the observation of the novel Q=0 magnetic state rises many questions. What can be the origin of this magnetism which preserves the translation symmetry of the lattice? Are we dealing orbital-like or spin moments ? Unfortunately, the information brought by the neutron experiment is rather scarce in terms of discriminating among models as, so far, the magnetic intensity has been only observed at three different Bragg peaks {\bf Q}=(1,0,L) (L=0,1 and 2) and is absent at {\bf Q}=(0,0,2) and {\bf Q}=(2,0,1). It is therefore delicate to determine a magnetic structure based on that. 

A way to distinguish among orbital and spin moments is to measure the magnetic form factor ($f(Q)$ in Eq. \ref{Imag}). Indeed, orbital-like magnetic moments, generated by current loops, are spatially extended objects and $f(Q)$ is expected to drop down at large $ \rm |Q|$ much fast that $f(Q)$ for spins. In the latter case, $f(Q)$ is given by Fourier transform of the electronic distribution around magnetic ions. Considering the small number of studied Bragg reflections, the interesting approach has turned to a dead-end. For instance, orbital and spin magnetic form factors would give no intensity or an extremely weak intensity at {\bf Q}=(2,0,1), as observed experimentally. To date, there is not enough data to tell whether the magnetic form factor is more consistent with orbital moments or spin moments.

Another way to get a deeper insight into the intrinsic nature of the novel magnetic order is to perform an analysis of the neutron intensity in terms of the magnetic group symmetry\cite{agterberg,arkady}. This analysis should be done relative to the crystalline group of cuprates. However, one additional difficulty is here that we even do not know which crystallographic sites can carry the magnetic moments. For instance, in orbital moments model, the moments do not  sit at given atomic positions. Further, the differentiation of models is as usual complicated due to the possibility of multiple domains. All of that implies that the symmetry analysis does not allow to discriminate different models, as the ones shown in Fig. \ref{Fig7}, based only on the limited neutron diffraction data although these models are clearly breaking different symmetries. 

One can nevertheless discuss the allowed symmetries as it has been done by Kaur and Agterberg \cite{agterberg} when considering the additional observation of dichroism in ARPES in Bi2212 \cite{kaminski} which, a priori, is related to the same broken symmetry phase. The observation of dichroism at the M=($\pi$,0) implies that both the fourfold rotation about z axis (C$_4$) and the diagonal mirror planes ($\sigma_{da}$, $\sigma_{db}$) are no longer symmetry operations in the pseudogap state. That reduces the number of possible magnetic point groups of the pseudogap phase to only two  \cite{agterberg}. Further, the photoemission matrix elements analysis of the dichroic signal \cite{simon} reduces the number of magnetic point groups to only one referred as $\underline{m}mm$, ${\rm E_u}$ irreductible representation of the ${\rm D_{4h}}$ point group \cite{agterberg}. The orbitals moments model created by current loops in the  $\rm CC-\Theta_{II}$ state (shown in Fig. \ref{Fig7}.A and FIG 2), proposed by Chandra Varma and coworkers\cite{simon,cmv06,arkady}, belongs to that magnetic group. 

%----------------------------------------------------------------------------------------------
\begin{figure}[t]
\centering
\includegraphics[width=15cm,angle=0]{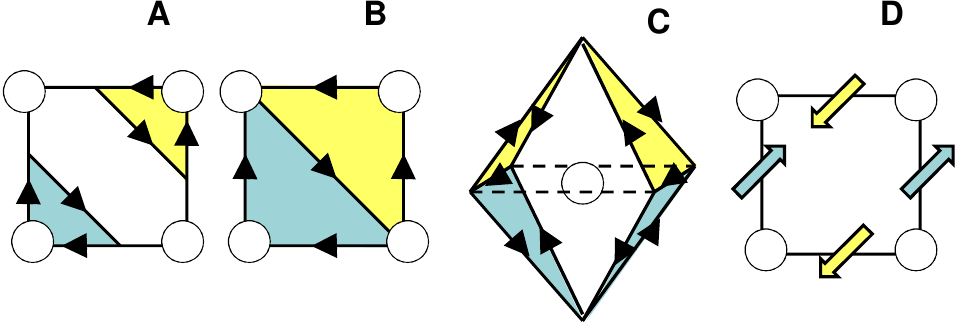}
\caption {
 Possible origins for the staggered magnetic pattern within a $\rm CuO_2$ plaquette with 3 orbitals moments configurations: A) $\rm CC-\Theta_{II}$ phase \cite{cmv06}, B) Two current loops filling the unit cell \cite{stanescu} and C) Delocalization of the circulation currents on the full CuO$_6$ octahedra \cite{weber}. D) Spins on oxygen atoms model. The direction of the moment could be in any direction as far as the moment on the oxygen along x is opposite to the one  on the oxygen along y. 
% . B) within the circulating current scenario, the orbital magnetic moments could be tilted with respect to the $\bf{c}$ in case of: (i) the 
% quantum superposition of the allowed 4 degenerated state of the $CC-Theta II$ phase, (ii) }
}
\label{Fig7}
\end{figure}
%----------------------------------------------------------------------------------------------

As a matter of fact, that $\rm CC- \Theta_{II}$ model gives magnetic intensity at the proper Bragg reflections. 
In contrast, our observation of a signal at the in-plane wave vector (1,0) or (0,1) dismisses the $\rm CC-\Theta_{I}$ phase (of 
${\rm B_{1g}}$ irreductible representation) with four current loops per unit cell \cite{cmv97,simon} as no signal is expected at these Bragg positions for that model. The same $\rm CC-\Theta_{I}$ phase was also ruled out \cite{simon} in Bi2212 by quantitative analysis of the ARPES-dichroism  \cite{kaminski}, only compatible with the $\rm CC- \Theta_{II}$ state. In principle, other magnetic structures could be also compatible with the observed magnetic intensity. One interesting example is a staggered orbital current phase proposed by Stanescu and Philipps \cite{stanescu} that we have reproduced in Fig. \ref{Fig7}.B. It has actually the same symmetry as the $ \rm CC-\Theta_{II}$ but does not explicit the role of oxygens atoms. As a result, it is also fully compatible with the magnetic neutron intensity observed around {\bf Q}=(1,0,L) Bragg spots and its absence at {\bf Q}=(0,0,L). These authors claim  that it is possible to generate such a translationally invariant current pattern within a one-band model \cite{stanescu} once oxygen orbitals are integrated out. The next nearest hopping, t', across the unit cell diagonal would be essential to get the pattern. Other staggered orbital current patterns are also compatible with the observed Bragg peaks once the apical oxygen orbitals are taken into account:  variational Monte-Carlo calculations in the extended Hubbard model indeed show that orbital moments develop on the copper oxygen octahedra \cite{weber}. Currents are flowing on certain faces of the CuO$_6$ octahedra giving rise to orbital moments tilted from $\bf{c}$ (Fig. \ref{Fig7}.C).

It is worth mentioning that spin based magnetic structure can also account for the neutron data. For instance, the model shown in Fig. \ref{Fig7}.D
\cite{fauque}  with spin moments on the in-plane oxygen atoms possesses the necessary ingredients: there is not net magnetic within the unit cell and the more intense magnetic Bragg peaks are for {\bf Q}=(1,0,L)$\equiv$(0,1,L). This order belongs to the ${\rm B_{1g}}$ irreductible representation and preserves spacial inversion contrary to the $ \rm CC-\Theta_{II}$ phase. 

\subsection{Magnetic excitations in Hg1201}

Having observed a long range magnetic ordered state, the next step is to search for the magnetic excitations associated
with that novel magnetic order. By principle, their observation would be an essential issue to validate and discriminate theoretical models. We have then performed inelastic polarized and unpolarized neutron experiments to look for magnetic excitations related to the pseudogap order. Recent measurements in Hg1201 reveal such a collective magnetic mode\cite{li2} around 54 meV. A peak is observed in the spin-flip channel and absent in the non-spin-flip channel in polarized neutron experiments. 

The mode is associated with the Q=0 magnetic order as the mode intensity is going to zero at the temperature, $\rm T_{mag}\equiv T*$ where the magnetic order vanishes and where anomalies occur in resistivity measurements. This has been observed for two different samples having very different T* in the underdoped regime.  Furthermore  the magnetic mode exhibits a very weak dispersion over the whole Brillouin zone. This behaviour drastically differs from the magnetic fluctuations usually reported in high-$\rm T_c$ cuprates \cite{cras} which occur systematically around the  planar antiferromagnetic (AF) wavevector {\bf Q}$_{AF}$=(0.5,0.5). Near the zone center, these AF fluctuations are absent. In contrast, the new excitations exist at any wave vector around nearly the same energy. That is what would be expected for an order parameter characterized by a discrete symmetry, such as in the case of Ising-like moments. 

This supports the orbital current phase, CC-$\rm \Theta_{II}$ \cite{cmv06}, which can be actually represented by an Ashkin-Teller model \cite{cmvaji}, characterized by two coupled Ising moments. More specifically, three weakly dispersive  modes are predicted from the $\rm CC-\Theta_{II}$ model \cite{cmvhe}. The two low energy modes are measurable by inelastic neutron scattering as they are related to a flip of an orbital moment by $\Delta$L=1. They can be classically represented by a rotation by $\pm$ 90$^{o}$, clockwise and anti-clockwise, from a loop current domain such as 
Fig. \ref{Fig7}.A. One of these modes is likely reported in \cite{li2}, whereas preliminary measurements indicate another mode at lower energy. 
The third mode corresponds to a flip of both moments in a domain like Fig. \ref{Fig7}.A: it then cannot be measured by neutron scattering as it is associated with a change of the orbital moment $\Delta$L=2. The experimental determination of that third excitation by other techniques like resonant inelastic X-ray scattering (RIXS) would be a crucial test for the theory. More generally, the observation of these weakly dispersive magnetic excitations challenge the theoretical description of high-$\rm T_c$ cuprates where magnetism essentially comes from AF correlation between Cu spins.
For the moment this observation of novel excitations is limited to Hg1201 system, it has to be generalized to other cuprates. 
 
\subsection{Direction of moments}
 
As shown above in section 3.1.3, the observed magnetic structure indicates a moment which does not point along a high symmetry direction but instead is tilted from the {\bf c} by $\phi \sim$ 45$^{o}$ (Fig. \ref{Fig5}.F). It is worth to mention that the Q=0 magnetism displays a large component along {\bf c}, contrary to usual Cu$^{2+}$ spin magnetism in undoped cuprates. That suggests a distinct origin. 

Clearly, the tilted picture is not consistent with the planar circulating currents models (Figs.~\ref{Fig7}.A and B) where the orbital magnetic moments have to be perpendicular to the current loops, {\it i.e.} pointing completely along {\bf c} axis. At first sight, it seems more natural for spins to explain a moment pointing in any direction.  However, this view might be too simple. Even in the CC picture, a few arguments can be given to explain the origin of the tilt. First, the proposed phase of circulating currents flowing on the face of ${\rm CuO_6}$ octahedra \cite{weber} obviously gives rise to a tilted moment with a proper order of magnitude. Second, even within planar orbital currents, corrections to the original CC-$\rm \Theta_{II}$ model can be presented. In YBCO, for instance, the Cu-O bond is not strictly perpendicular to {\bf c} as copper and oxygen atoms are not in the same plane, yielding a tilt of about 14$^{o}$. Still in YBCO, an in-plane spin component necessarily occurs due to spin-orbit coupling \cite{cmvvivek} whose amplitude might be measurable with neutrons although it would not be a very large contribution. In Hg1201, these two last points are not present due to its simple tetragonal lattice symmetry. Nevertheless, even for simple squared lattice, an in-plane component can appear in the CC-$\rm \Theta_{II}$ model due to quantum corrections of the ground state \cite{cmvhe}. Indeed, the $\rm CC- \Theta_{II}$ is  characterized by 4 distinct states for a given $\rm CuO_2$ plaquette. If each state is considered independently from the others, that gives 4 classical domains. Alternatively, once quantum corrections are introduced, the ground of the $\rm CC- \Theta_{II}$  becomes a quantum superposition of these 4 states. In neutron diffraction measurements, what is interpreted in terms of a planar component  could simply result from an interference phenomenon: in such a case, the tilt angle indicates the degree of admixture of the 4 CC states within the ground state. Within this approach,  the exact amount of the tilt is still under question, as it is directly related to te dispersion of the new Q=0 magnetic modes\cite{cmvhe}. 

In addition to other symmetries discussed above, it is interesting to note that the existence of the in-plane component breaks the $\sigma_z$ mirror plane. In Landau theory of continuous second order phase transition, this indicates that each magnetic component is represented by two different order parameters likely having distinct symmetries. This means in principle that both order parameters would also exhibit different critical temperatures. 

\subsection{Nuclear Magnetic Resonance and muon-spin rotation: silent local probes}

Besides circulating currents, we have mentioned that the  Q=0 AFO could also be produced by spins located on oxygen sites. There is however a noticeable difference between a spin-based model and an orbital-based model.
Spins located on oxygen sites should produce a sizeable broadening of the Nuclear Magnetic Resonance (NMR) $^{17}$O lines which is not observed \cite{bobroff1}. At variance, the staggered orbital moment of CC phases \cite{cmv97,simon,cmv06,stanescu}  would  not give rise to any measurable signal in NMR experiments. By symmetry, the effects of orbital-like moments cancel out for each nucleus site. As a matter of fact, there is no indication in NMR measurements  of  a static magnetic ordering at T*. In underdoped YBCO, there is for instance no broadening of the $^{89}$Y NMR line which actually sharpens up at low temperature \cite{bobroff2}. Recently, a $^{89}$Y NMR study has been misleadlingly used to point out a lack of evidence for orbital-current effects in the cuprate ${\rm Y_2Ba_4Cu_7O_{15-\delta}}$ \cite{strassle}. This system contains 2 adjacent $\rm CuO_2$ planes with different hole doping of about $\sim$ 3 \% due to different environment. Owing to this charge imbalance, the authors of Ref.~\cite{strassle}  expected a finite broadening of the $^{89}$Y NMR line for orbital currents phase and they did not observe it. Unfortunately, this expectation is not realistic for all the CC phases respecting the lattice invariance symmetry discussed here \cite{cmv97,simon,cmv06,stanescu} because the magnetic field at the $^{89}$Y site almost cancels out the effect as there are a few opposite current loops around each yttrium nucleus.

In principle, zero-field muon-spin rotation ($\mu$SR) technique is a very powerful to detect very tiny magnetic moments.
A signature of the Q=0 AFO reported by polarized neutron should be visible in $\mu$SR measurements.
Unfortunately, interpretation of $\mu$SR results have been quite contradictory on that topic over the last decade 
\cite{sonier1,sonier2,sonier,uemura}. 
At first, small spontaneous static magnetic fields of electronic origin were reported in YBCO to be intimately related to the pseudogap transition \cite{sonier1}. Later on, the same data were interpreted as being due to spatial charge inhomogeneities in relation to the CuO chains \cite{sonier2}. Next, a magnetic order was observed but did not evolve significantly with hole doping \cite{sonier}. In only one sample, a second magnetic component was interpreted as due to dilute impurity phases \cite{sonier}. That signal was misleadlingly associated with the magnetic order that we reported in neutron diffraction \cite{mook} which in contrast to $\mu$SR is observed systematically in all underdoped YBCO samples that have been studied \cite{berlin,fauque}. The  underdoped YBCO sample  carrying a dilute impurity phases in the ($\mu$SR) study \cite{sonier} is known to exhibit a magnetic signal around {\bf Q}$_{AF}$=(0.5,0.5), easily detected by neutron diffraction \cite{mookpp}. It is notorious that such a kind of parasitic AF phase does not exist in all underdoped YBCO samples \cite{stock}. It then appears natural to ascribe the $\mu$SR impurity phase  to the parasitic AF signal contrary to what has been claimed in \cite{sonier}. Another $\mu$SR in LSCO came to the conclusion of the absence of broken time-reversal symmetry in the pseudogap state \cite{uemura}. All these reports then suggest that $\mu$SR probe is silent with regard to the magnetic order reported in polarized neutron diffraction as it is also the case for NMR. 

Actually, two independent theoretical calculations \cite{cmvmuons,millismuons} estimate the screening charge density distribution due to a point charge, such as that of a positive muon $\mu^+$, placed between the planes of a highly anisotropic layered metal. Clearly, a muon is shown to lead to an observable local perturbation.  This questions the interpretation of muon-spin-rotation experiments in metallic high-temperature superconductors. For the Q=0 AFO, the field at the muon site was expected to be of several mT according to the neutron data. This field is actually
reduced by more than 2 orders of magnitude due to muon related screening \cite{cmvmuons}. That seems to resolve the apparent contradiction between the polarized neutron diffraction and $\mu$SR experiments.

 \subsection{Electronic liquid crystal}

%----------------------------------------------------------------------------------------------
\begin{figure}[t]
\centering
\includegraphics[width=12cm,angle=0]{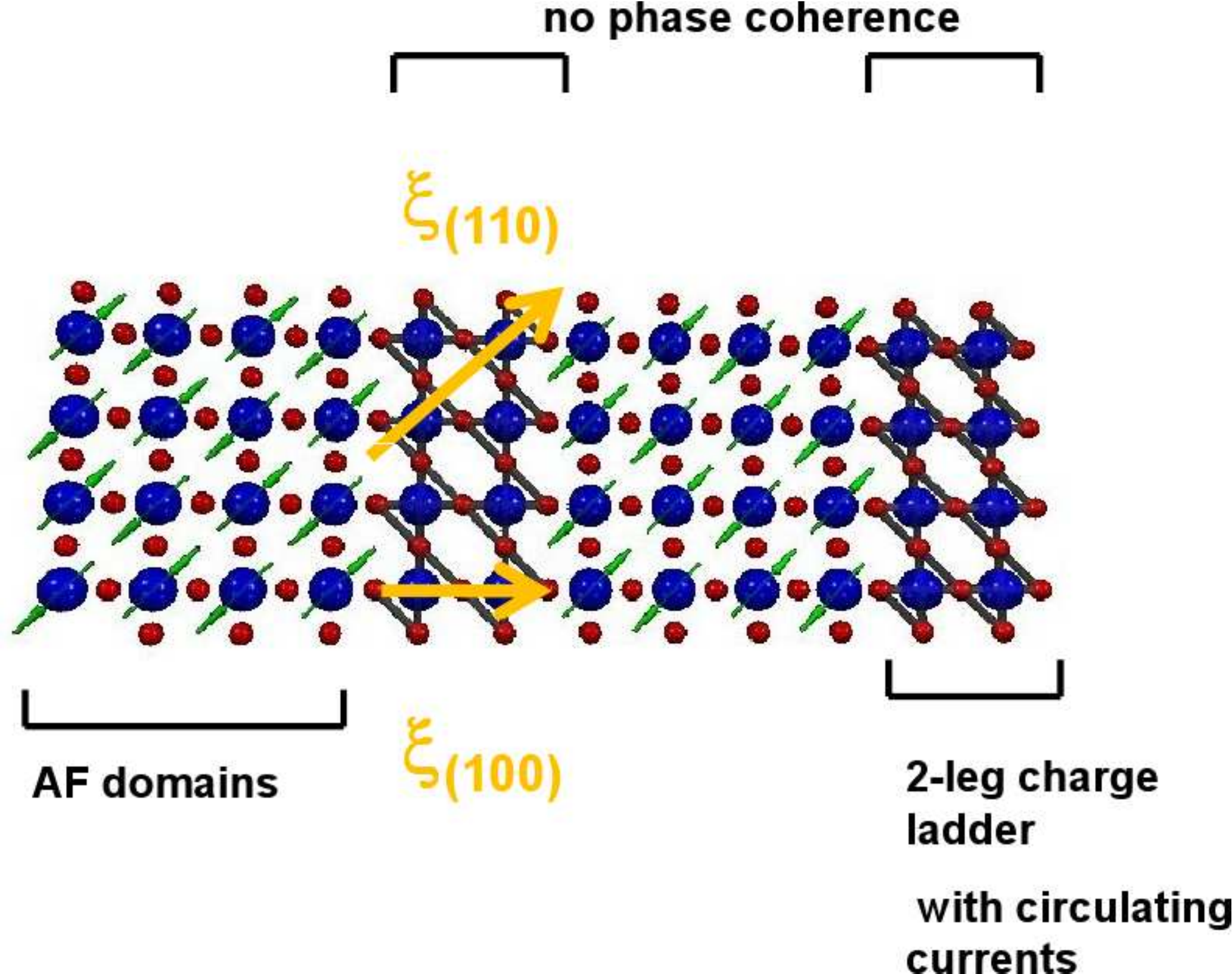}
\caption {
Cartoon picture suggesting the indication of a spin and charge stripe structure and an orbital-like magnetic order. The stripe structure is characterized by doped holes located on two-leg ladders, separating AF domains. The charge periodicity is $\sim 6 a$ and the magnetic periodicity is $\sim 12 a$. 
}
\label{Fig12}
\end{figure}
%----------------------------------------------------------------------------------------------

It has been recently proposed that a quantum analog of liquid crystal states - electronic smectic and nematic phases - may play an important role in the physics of strongly correlated systems \cite{Fradkin}, such as high temperature superconductors. The smectic and nematic phases are characterized by the broken unidirectional translational
and rotational symmetry and broken rotational symmetry, respectively. In cuprates,  the electronic liquid crystal (ELC) phases may have different origins.   Starting from a hole doped AF Mott insulator,   electronic smectic or nematic phase could correspond to static or fluctuating stripes \cite{stripes,vojta}.  They can also correspond to static or fluctuating spin spiral phases \cite{Sushkov}. Starting from the metallic side of cuprates phase diagram, an electronic nematic phase can take the form of a spontaneous distortion of the Fermi surface \cite{Pomeranchuck}, breaking the a-b symmetry.

Direct or indirect evidence of electronic smectic and nematic phases have been reported in superconducting cuprates, via different experimental techniques \cite{Fradkin}. For instance, in strongly underdoped YBCO$_{6.45}$ \cite{hinkov},   quasi-1D incommensurate (IC) spin fluctuations  at  {\bf Q}$_{IC}$=$(0.5\pm \delta,0.5)$  spontaneously develop  below  $\sim$ 150 K and become static at very low  temperature. The observation can be generalized for hole doping below 9\% (corresponding to an oxygen content $<$ 0.5 in YBCO$_{6+x}$) \cite{haug1}. In strongly underdoped YBCO, both neutron scattering \cite{hinkov,haug1} and resistivity \cite{Ando-PRL02} measurements indicate that the a-b symmetry is broken. These observations strongly support the existence of ELC phases in strongly underdoped YBCO, but there is not yet a consensus concerning the electronic instability at the origin of the phenomenon. In twinned LSCO samples, IC spin fluctuations at {\bf Q}$_{IC}$=$(0.5\pm \delta,0.5)\equiv (0.5, 0.5\pm \delta)$ are  also observed and are usually interpreted in terms of fluctuations stripes oriented along Cu-O bonds (either along {\bf a} or {\bf b} due to the existence of twin domains). The nematic ELC phase is therefore associated with the appearance of these fluctuating stripes, breaking the a-b symmetry of the system \cite{stripes,Tranquada-review}. Obviously, these ELC phases and the novel Q=0 magnetic order correspond to  different electronic instabilities. This leads a simple question: what happens when these phases met in the phase diagram of cuprates ? 

To address this issue, it is first interesting to consider the case of LSCO where we have reported a short range Q=0 AFO\cite{baledent}.  Interestingly, low energy IC spin fluctuations display a feedback to the short range ordering. The incommensurability parameter $\delta$ is renormalized at $\rm T_{mag}$. The original picture of stripes physics implies that the doped holes self-organise into lines of charged stripes which create antiphase domain walls between antiferromagnetic domains where Cu$^{2+}$ spins fluctuate and freeze at low enough temperature. A sketch of such charge and spin organisation is given in Fig.~\ref{Fig12} for a hole doping of 0.085 $\sim$ 1/12 corresponding to our LSCO sample \cite{baledent}. We here consider the case of bond-centered stripes where a doped hole is located on two-leg ladders. The orbital-like order could be confined within the two-leg ladders, yielding typical correlation lengths of $\sim 2a$ along the (100) direction and $\sim 3 a$ along the (110) direction, as observed experimentally (see Figs. \ref{Fig10}.A and B). The orbital-like magnetic order should not develop a phase coherent from one ladder to another, otherwise the magnetic intensity should appear at incommensurate wave vectors away from H=1, instead of the broad peak observed around $\bf{Q}$=(1,0,L). This example in LSCO underlines the competition between ELC and orbital-like Q=0 instabilities.

In YBCO, neutron scattering measurements provides evidence of ELC phases, but only in the strongly underdoped regime (p$<0.09$).
The Q=0 AFO has been studied  from $\rm YBCO_{6.5}$ (p=0.09) up to $\rm (Y,Ca)BCO_7$ (p=0.2). In this hole doping range, neutron scattering studies do not indicate that the ELC phases dominate the low energy properties of the system. In $\rm YBCO_{6.6}$ (p=0.12) for instance, the spin excitation spectrum exhibits a spin pseudogap in the normal state and low energy spin fluctuations are hardly measurable. Above the spin pseudogap, IC spin fluctuations still display an a-b anisotropy in the energy range between $\sim$20 meV and $\sim$38 meV \cite{hinkov1}. To date, one still ignore at which temperature this anisotropy appears.  In the hole doping range where the Q=0 AFO has been identified, an a-b anisotropy  has been recently detected in the Nernst effect \cite{daou}. In principle, this supports theoretical approaches \cite{stripes,vojta} in which the C$_4$ symmetry is spontaneously broken in the pseudogap state, owing to strong electronic correlations yielding the nematic ELC state. This anisotropy develops when entering the pseudogap state at a temperature corresponding exactly to $\rm T_{mag}$ of the Q=0 order in a large doping range. Clearly, the anisotropy starts at a temperature where there is no indication of a similar anisotropy neither in the spin excitation spectrum nor in electrical resistivity. In addition, the existence if an a-b anisotropy in the Nernst effect has not been studied in strongly underdoped YBCO, where the ELC phases take place. Actually, it exists a completely different interpretation of this anisotropic Nernst effect \cite{cmvyakokapil}. In this interpretation, the anisotropic Nernst effect would appear as a consequence of the time-reversal symmetry breaking below T* in agreement with polarized neutron data  \cite{fauque,mook} and Kerr effect \cite{xia}. 

In  underdoped YBCO for p$>$0.09, quantum oscillations data indicate the existence of small electron Fermi pockets \cite{quantum-osc,proust,sebastian} at very large magnetic field (H$>$40 Tesla) when superconductivity is supposed to be destroyed. These data  are often interpreted in terms stripe-like spin density wave \cite{Fsurf-recons}. Recent transport \cite{proust2010} and quantum oscillations \cite{singleton} data further indicate that the existence of these electron Fermi pockets is quite sharply peaked around the hole doping of 1/8, where stripes are expected to be pinned down by the lattice. One can be tempted to establish a connection between the ELC phase reported in strongly underdoped YBCO (for p$<$0.09) and the  small Fermi pockets reported at larger doping (p$>0.09$) and very high magnetic field (H$>$40 T). This interplay is further supported by the enhancement of the smectic ELC phase in $\rm YBCO_{6.45}$ when applying a modest magnetic field of 14 Tesla \cite{haug}.  Strictly speaking,  however, there is no compelling evidence of well developed ELC phases in the hole doping range where the quantum oscillations are reported, questioning this straightforward scenario. 

At variance with LSCO,  in YBCO there is no clear indication that the Q=0 AFO and ELC phases could be found in the same portions of  the phase diagram versus hole doping and magnetic field. In order to get better a understanding of the interplay between these phases, the search for the Q=0 magnetic order should be extended to the strongly underdoped regime. The search for such an order at high magnetic field is for the moment technically impossible with polarized neutron diffraction technique. However, instead of applying a uniform perturbation to the system, one should consider the effect of  local perturbations induced by the substitution of non magnetic Zn or magnetic Ni impurities. If there were a competition between the Q=0 magnetic order and the ELC phase, their balance could be modified around local defects.

\section{Conclusion}

The elucidation of the pseudogap phenomenon of the cuprates has been a major challenge in condensed matter physics for the past two decades. Following initial indications of broken time-reversal symmetry in photoemission experiments\cite{kaminski}, recent polarized neutron diffraction works demonstrated the universal existence of an unusual magnetic order below T* in 3 different cuprates families \cite{fauque,mook,li1,baledent}
whatever the number of CuO$_2$ plane per unit cell. The observed magnetic order can be interpreted as a Q=0 antiferromagnetic order preserving the translational symmetry of the lattice. The observed symmetry  is  furthermore consistent with a particular type of order involving circulating orbital currents  ($\rm CC-\Theta_{II}$ phase) \cite{simon,cmv06}, and with the notion that the phase diagram is controlled by a quantum critical point  \cite{cmvvivek}. The ordered temperature indeed extrapolates  to zero  at  the  doping  level  of  p $\sim$ 0.19  corresponding to the end point if the pseudogap state. To date, it is the first direct evidence of a hidden order parameter associated with the pseudogap phase of high-$\rm T_c$ cuprates. It has the profound implication that the pseudogap transition T* is a genuine phase transition although no sharp thermodynamic anomalies are observed \cite{varmaQCP}. Uniform susceptibility \cite{leridon} and magneto-optic Kerr effect \cite{xia} measurements confirm this picture.  Recent measurements in ${\rm La_{2-x}Sr_xCuO_4}$ reveal that a similar ordering occurs in that system for x=0.085 \cite{baledent}. However, the observed order is  at short range and  bidimensional.  That difference might be related to the competition of the Q=0 magnetic order with the stronger tendency towards an electronic liquid crystal state in that system.

% The Acknowledgements are also a un-numbered section
\section*{Acknowledgements}
We particularly acknowledge Victor Bal\'edent, Benoit Fauqu\'e, Martin Greven, Yuan Li and all our collaborators whose names appear in references \cite{fauque,mook,li1,baledent,li2}. We wish to thank Vivek Aji, Henri Alloul, Marc Gabay, Thierry Giamarchi, Marc-Henri Julien, Brigitte Leridon, Louis-Pierre Regnault, Arkady Shekhter and Chandra Varma for intense discussions on various aspects related to this work.

\end{document}